\documentclass[lettersize,journal]{IEEEtran}

\usepackage[noadjust]{cite}
\usepackage{amsmath,amssymb,amsfonts}
\usepackage{glossaries}

\usepackage{algorithm}

\usepackage{graphicx}
\usepackage{textcomp}
\usepackage{xcolor}
\usepackage[capitalise]{cleveref}
\usepackage{siunitx}
\usepackage[export]{adjustbox}
\usepackage[caption=false,font=normalsize,labelfont=sf,textfont=sf]{subfig}
\usepackage{tabularx}
\usepackage{algpseudocode}
\usepackage{xurl}
\usepackage{flushend}
\usepackage{tablefootnote}
\usepackage[raggedrightboxes]{ragged2e}
\usepackage{graphicx}

\crefname{subsection}{Subsection}{Subsections}

\DeclareSIUnit\operations{op}
\usepackage[colorinlistoftodos,prependcaption,textsize=tiny]{todonotes}

\def\BibTeX{{\rm B\kern-.05em{\sc i\kern-.025em b}\kern-.08em
    T\kern-.1667em\lower.7ex\hbox{E}\kern-.125emX}}

\newacronym{scd}{SCF}{Spectral Correlation Function}
\newacronym{fft}{FFT}{Fast Fourier Transform}
\newacronym{coh}{COH}{Spectral Coherence Function}
\newacronym{cw}{CW}{Continuous Wave}
\newacronym{fam}{FAM}{Fourier Accumulation Method}
\newacronym{snr}{SNR}{Signal to Noise Ratio}
\newacronym{bpsk}{BPSK}{Binary Phase Shift Key}

\newacronym{ppd}{PPD}{Privacy Protection Device}

\newacronym{gnss}{GNSS}{Global Navigation Satellite System}
\newacronym{gps}{GPS}{Global Positioning System}
\newacronym{nmea}{NMEA}{National Marine Electronics Association}
\newacronym{nasa}{NASA}{National Aeronautics and Space Administration}
\newacronym{agps}{A-GPS}{Assisted/Augmented GPS}
\newacronym{scer}{SCER}{Security Code Estimation and Replay}
\newacronym{nma}{OS-NMA}{Open Service Navigation Message Authentication}
\newacronym{sis}{SIS}{Signal In Space}
\newacronym{raim}{RAIM}{Receiver autonomous integrity monitoring}
\newacronym{pnt}{PNT}{Position-Navigation-Time}
\newacronym{rtk}{RTK}{Real-time Kinematics}
\newacronym{agc}{AGC}{Automatic Gain Control}
\newacronym{meo}{MEO}{Medium Earth Orbit}
\newacronym{rssi}{RSSI}{Received Signal Strength Indicator}
\newacronym{llh}{LLH}{Latitude, Longitude and Height}
\newacronym{sqm}{SQM}{Signal Quality Monitoring}

\newacronym{sdr}{SDR}{Software Defined Radio}
\newacronym{sdrs}{SDR}{Software Defined Radios}
\newacronym{ots}{OTS}{Off-the-Shelf}
\newacronym{rt}{RT}{Real-Time}
\newacronym{ide}{IDE}{Integrated Development Environment}
\newacronym{fifo}{FIFO}{First-In-First-Out}
\newacronym{sma}{SMA}{SubMiniature Version A}
\newacronym{rf}{RF}{Radio Frequency}
\newacronym{uav}{UAV}{Unmanned Aerial Vehicle}
\newacronym{vtol}{VTOL}{Vertical Take Off and Landing}
\newacronym{ota}{OTA}{Over-The-Air}
\newacronym{ism}{ISM}{Industrial Scientific and Medical}

\begin{document}

\title{GNSS jammer localization and identification with airborne commercial GNSS receivers}

\author{
Marco Spanghero,
Filip Geib,
Ronny Panier,
Panos Papadimitratos,~\IEEEmembership{Fellow,~IEEE}

\thanks{
This work was funded in parts by the Swedish Research Council, the Swedish Foundation for Strategic Research, and the Strategic Research Area on Security and Emergency Preparedness.\\
Corresponding author: Marco Spanghero\\
Marco Spanghero and Panos Papadimitratos are with the Networked Systems Security Group, KTH Royal Institute of Technology, 164 40, Stockholm, Sweden. 
(e-mail: marcosp@kth.se, papadim@kth.se). Filip Geib and Ronny Panier are with Wingtra AG, Zürich, Switzerland. 
(e-mail: filip.geib@wingtra.com, ronny.panier@wingtra.com).}
}

\maketitle

\begin{abstract}
Global Navigation Satellite Systems (GNSS) are fundamental in ubiquitously providing position and time to a wide gamut of systems. Jamming remains a realistic threat in many deployment settings, civilian and tactical. Specifically, in Unmanned Aerial Vehicles (UAVs) sustained denial raises safety critical concerns. This work presents a strategy that allows detection, localization, and classification both in the frequency and time domain of interference signals harmful to navigation. A high-performance Vertical Take Off and Landing (VTOL) UAV with a single antenna and a commercial GNSS receiver is used to geolocate and characterize RF emitters at long range, to infer the navigation impairment. Raw IQ baseband snapshots from the GNSS receiver make the application of spectral correlation methods possible without extra software-defined radio payload, paving the way to spectrum identification and monitoring in airborne platforms, aiming at RF situational awareness.
Live testing at Jammertest, in Norway, with portable, commercially available GNSS multi-band jammers demonstrates the ability to detect, localize, and characterize harmful interference. Our system pinpointed the position with an error of a few meters of the transmitter and the extent of the affected area at long range, without entering the denied zone. Additionally, further spectral content extraction is used to accurately identify the jammer frequency, bandwidth, and modulation scheme based on spectral correlation techniques.

\end{abstract}

\section{Introduction}

\gls{gnss} provide precise and available navigation and timing to a wide range of platforms, from industrial systems to autonomous vehicles. Due to the distant orbital planes of \gls{gnss} constellations which are situated in \gls{meo}, the received signal strength on the ground is in the order of \SI{-130}{\dB m} on average, for GPS L1 signals. This makes the majority of commercially available \gls{gnss} receivers vulnerable to interference, specifically intentional ones like jamming. The current global situation accentuated the importance of this issue, not only from a tactical perspective but also in the civilian segment. While large and powerful tactical jammers are indeed a problem, the majority of jamming events in the civilian sector 
are due to the great availability of low-cost, portable, and effective interference transmitters. Jamming is often used to disable \gls{gnss} receivers used in tracking goods and people, geofencing of restricted areas, or law enforcement. Several examples of such behavior exist \cite{SkytruthJamming,CNETNewark,insideGNSSNewark}, making jammer localization an important topic.

As an attack vector, jamming is as simple as effective: it can deny or severely degrade the victim receiver a valid \gls{pnt} solution and it can consistently be used as a stepping stone to mount more sophisticated attacks, such as spoofing \cite{HumphreysAssessingSpoofer,Humphreys2012,HuangL2015} and meaconing \cite{LenhartSP:C:2022,10.1145/3558482.3590186}. A short jamming phase is often functional for capturing a victim device whose takeover location is unknown to the adversarial transmitter or when the victim is moving.

%The aim is dual: to make \gls{gnss} receiver robust to jamming and interference in general, and localize the adversarial transmitter while identifying what type of signal the transmitter is broadcasting. The latter allows estimating the effects it has on the receiver's \gls{pnt}. 

In particular, for \gls{uav} intentional or unintentional interference poses a significant security risk as a degraded positioning solution can only partially be addressed by other forms of navigation in consumer devices. Intuitively, a \gls{uav} would try to avoid operation within a denied area: this is important for operational safety, but it requires identification, localization, and mapping of the interference source. Additionally, this needs to be performed at long range, outside the affected area. By achieving such, the system can be made overall more robust to interference and active measurements can be taken toward removing the interference source. 

In the field, localization is often neither practical nor efficient, specifically when the survey area is large or hard to explore due to the terrain or other constraints. Additionally, front-end level enhancements allow receivers to adaptively recover from certain types of jammers, but such countermeasures are often reactive, meaning that the receiver does estimate in advance, based on the spectral structure of the jammer, if the interference effect can be mitigated. 

% Existing methods relying on \gls{sdr} spectrum sensing add considerable complexity to the hardware of the receiver and are not normally available in consumer platforms. Furthermore, jamming and spoofing are a significant threat to autonomous aerial vehicles, and sensing, localization, and avoidance capabilities can largely improve the robustness of such systems. The combination of advanced flight dynamics with spectrum sensing capabilities enhanced early detection and awareness of problematic areas, where the \gls{gnss} conditions might be impaired. 

Overall, while the problem of jamming detection is well understood and researched \cite{7444136} the localization and identification of the interference source over large areas is a complex problem that often requires specialized tools and knowledge, like dedicated interference finding equipment \cite{Anritsu_appnote_rfi}. Recent developments for \gls{vtol}s allowing high maneuverability and longer flight times make them attractive to map large areas. By a clever combination of flight dynamics and a commercial onboard \gls{gnss} receiver and antenna orientation, it is possible to streamline RF spectrum survey operations. Even in the presence of adversarial transmitters, effective localization, and mapping of the adversary can be achieved without entering the denied zone.

Preliminary experimentation based on radio amateur equipment causing interference shows that airborne localization of transmitters is both effective and accurate, but cannot distinguish different types of transmitters and estimate the degradation of the \gls{pnt} solution in the affected area \cite{SpangheroMGPP:C:2024}. 
Here, we improve the detection algorithm by extending the localization of the jammer with the classification of the spectral structure based on snapshots provided by a commercial \gls{gnss} receiver instead of the general approach using \gls{sdr}. Practically, our solution eliminates the need for customized additional hardware. 

First, we thoroughly test and validate our approach with multi-constellation, and multi-frequency \gls{gnss} jammers in a controlled environment. Then, we present the results of flight missions performed at Jammertest 2023 \gls{gnss} testing campaign, evaluating the proposed approach against live transmitters \cite{jammertest-event}. We assess the interference detection, localization, and spectrum sensing capabilities in a real context, significantly extending the initial results in \cite{SpangheroMGPP:C:2024}, with jammers transmitting over the air, focusing on personal privacy protection that is available in the open market, generally known as \gls{ppd}.  The system performs correct localization in the outdoor measurement cases and identification of different transmitters in both the laboratory and field tests, concretely showing that \gls{vtol}s can be relied upon for RF spectrum assessment even in adversarial conditions.

The rest of the work unfolds in the following way: \cref{sec:related_work} discusses the relevant related work in the open literature, \cref{sec:adv-sys} shows the system and adversary model while \cref{section:methodology} presents the approach used in this work. \cref{sec:experiment_setup} discuss the system model and the experimental setups used to validate the results, \cref{sec:evaluation} discusses validation experiments conducted in a controlled environment and live testing at Jammertest 2023. Finally, \cref{sec:conclusion} concludes the work and discusses future developments.

\section{Related Work}
\label{sec:related_work}

Jamming detection is a well-known problem. Methods based on monitoring abnormal power density in the \gls{gnss} frequency bands can reliably detect the presence of potentially adversarial transmitters \cite{WOS:000377112000007} but often do not provide any information on the exact nature of the transmission (practically, the detector is not aware of the signal structure of the adversarial transmitter). Methods that monitor the overall in-band power relevant to the \gls{gnss} spectrum estimate the received power density level by observation of the \gls{gnss} receiver Automatic Gain Control (\gls{agc}) and amplifier stage gain \cite{WOS:000209006500003,WOS:000874785702002,WOS:000375213002049} or rely upon direct IQ sample manipulation \cite{WOS:000423143700015,WOS:000429990900016,WOS:000414280900018,WOS:000359380700146} (e.g. methods based on spectrum analyzers or software-defined radios). 

Monitoring the receiver's Automatic Gain Control (\gls{agc}) provides a good estimation of the received signal power. The \gls{agc} controls the front-end gain to maintain full dynamics at the analog-to-digital converter. High-interference environments will cause the \gls{agc} value to be low, and vice-versa \cite{akos2003}. On the other hand, it is complex to relate the actual \gls{agc} variations to adversarial action or changes in the environment. Specifically, multipath and environment geometry-induced signal quality variations are known to cause changes in the \gls{agc} that cannot be related to any adversarial action (e.g., entering or exiting a tunnel), and overall \gls{agc} calibration is complex due to differences between the implementation in the receiver hardware \cite{WOS:000209006500003}. \gls{agc} variations (and similarly other signal quality indicators) provide a single-value representation of the overall noise level in a particular channel of the front end and for this reason, albeit effective and low-cost to evaluate, they cannot provide a precise understanding of the radio frequency (RF) spectrum environment situation. 

% To be seen where to fit it, nowhere good atm
Cooperative schemes where multiple sensors are deployed in the field (e.g., modern mobile phones with \gls{gnss} raw measurement capabilities) extend the capabilities in detecting interference beyond the range of a single sensor \cite{BorioCoordinates2016,WOS:000940628000001,WOS:000327163303020,ENC2023-15441}. This can be achieved with a network of dedicated fixed sensors \cite{WOS:000333247400024,WOS:000436563900049,WOS:001022344800060} or mobile phone crowd-sourced measurements \cite{WOS:001022344800062,OlssonALP:C:2022}. Two major issues arise in this context. First, dedicated fixed networks of sensors are expensive to deploy and do not provide any flexibility if the area to be monitored changes or new areas of interest emerge. Second, the estimated position of the interference depends on the quality and trustworthiness of the participant-provided measurements. While the first issue is design-dependent, the second can be mitigated with the adoption of security-focused participatory sensing schemes \cite{GisdakisGP:J:2016,GisdakisGP:C:2015}.

Exploiting direct baseband sampling of the \gls{gnss} spectrum provides much more information at a significant computational cost. High sampling rate and bandwidth requirements of \gls{sdr}s make processing of such recordings challenging if not impossible on mobile devices. To more extensively address the challenge of jamming detection, a few high-end consumer receivers are starting to provide the user with power spectral density measurements (notably Septentrio \cite{mosaicx5} and U-Blox \cite{ubloxf9p}) but to the best of our knowledge such measurements are only an indication of the power density rather than a calibrated value. The u-Blox F9 family of receivers provides the user with an internally calculated power spectral density, while the Septentrio counterpart allows access to the raw baseband samples but at a pre-determined rate, lower than the one internal to the receiver. 

These samples are a snapshot of the current \gls{gnss} front-end measurements and provide a higher level of information, especially in the frequency domain, and are a great trade-off between the performance of \gls{sdr} baseband sampling and signal quality monitoring \cite{WOS:000375211601043}. Power-time and frequency-time analysis leverage the transmission pattern of jammers and general interference sources to detect spurious emissions, classify them, and categorize them based on their properties. Power-time countermeasures consider time-series variation of the received signal strength to identify behaviors consistent with an adversarial transmitter \cite{WOS:000403758600007,WOS:000318940000010,Borio2016,van_der_Merwe_2023}. 

Machine learning aided techniques require training datasets that are difficult to generate, generalize, and overall complex to maintain \cite{278b52b90206426bae50f2f582b4aeb1,WOS:000597751100006,WOS:000850372300015}. Template matching of unknown signals leads to poor classification performance and generally limits the number of identifiable signals to the already known classes. 
For this reason, model-free signal identification has a more flexible approach. A blind recognition approach is suitable for analyzing signals that present cyclostationary features (e.g., jammers whose statistical properties repeat over time) and pulse-based interference signals \cite{WOS:000568618903045,WOS:000377112000013,WOS:000305396000003}. Recent literature shows that detection and geo-localization of both are possible even from space, but such techniques are limited by access to a constellation of low-orbit satellites \cite{WOS:001022344800055}.

Regarding Earth-bound measurements, different methods rely on cyclostationary properties of the signals rather than just signal power and exhibit a much higher detection accuracy and true positive rate. With high-resolution spectrum measurements, it is possible to identify the type of transmitter and the frequency pattern but was only demonstrated on ground-based \gls{sdr} measurements \cite{WOS:000354717200013,WOS:000290749900023,WOS:000388865300019,WOS:000424862200034}. Although similar approaches work for various mobile applications (e.g., moving vehicles in \cite{WOS:000464724000001}), such an approach was not shown in \gls{uav} in combination with transmitter location finding. 

In the context of airborne platforms, jammers are a significant threat as they effectively deny precise locations used in trajectory estimation. While inertial navigation is possible for reasonably short time intervals, \gls{vtol}s largely depend on \gls{gnss} for location and navigation. As a form of simple but effective attack, jamming is often preferred over spoofing for aerial devices, as it can be equally effective at a much reduced complexity

Localization of interference transmitters based on airborne platforms was demonstrated in the \gls{ism} band, localizing a Wi-Fi router and using an equivalent platform in open-air testing for localization of a wideband emitter. Examples in open literature rely on multi-rotor platforms and can locate GPS jammers \cite{WOS:000365040300004}. While such platforms can navigate for a short time (about \SI{20}{\minute} of total flight time) and successfully position even in denied conditions, there is always a risk connected to operations based within denied areas, specifically if the interference is due to spoofing, rather than jamming. Similarly, cooperative autonomous detection of jammers based on \gls{vtol}s allows finding and locating the interference transmitter \cite{9438441}, but results in open literature often do not include field observation. Additionally, the work in \cite{9438441} assumes that \gls{vtol}s can reliably fly within the degraded zone, and given a powerful enough jammer that might pose a security risk. The benefits of using fixed-wing planes with \gls{vtol} capabilities are shown in \cite{SpangheroMGPP:C:2024}, where the flight time is significantly extended and the \gls{uav} performs long-distance detection without entering the denied zone. Additionally, few examples are available where the sampling front-end is not a dedicated antenna and radio platform and relies on representative frequency information from the \gls{gnss} receiver frontend allowing improving existing methods based on signal power monitoring in deployed mobile platforms.

\section{System and Adversary model}
\label{sec:adv-sys}
In this work, we consider an airborne platform capable of both fixed-wing flight and hovering (e.g., maintaining a certain position) with the ability to efficiently transition, without landing between the two modes. The \gls{gnss} receiver in the platform is provided with a semi-directional antenna (e.g., whose radiation pattern is narrower than hemispherical in the direction of the main lobe) mounted on the top of the vehicle. The receiver provides RF spectrum snapshots at all frequencies with programmable rate, but discrete in time. Additionally, the vehicle can semi-autonomously fly on a pre-determined path with transitions between fixed-wing flight and hovering that are triggered by the operator on the ground. Generally, one would want to perform localization of the impaired areas without being inside a denied zone: leveraging the gain and directionality of the antenna, this can be done from afar. The vehicle performs multiple scanning operations at different locations when in hovering mode, providing spatially separated data.

The adversary is a ground-based transmitter whose objective is to degrade the quality of the \gls{gnss} reception so that the receiver loses lock on the legitimate signals. The jammer, while portable and mobile, is supposed to be mostly static during its operation. On the other hand, we make no restrictions on the behavior the jammer has in terms of frequency and power, with time-varying output levels and transmitted waveforms. Suppose the jammer is present and active when the receiver is activated and in cold start. Such a case is simpler from an adversarial perspective as often the sensitivity in the acquisition phase in a commercial receiver is lower than tracking an acquired signal, making the receiver more vulnerable. Nevertheless, strong enough interference will cause a loss of lock even when the receiver is fully tracking. It is also possible for a receiver to still produce a viable \gls{pnt} solution during jamming (i.e. some selective jammers might only disturb specific constellations, leaving others viable) but still with a considerable degradation.

Notably, the transmission of interference has a very unbalanced nature in the effort required by the attacker. This makes the attack easy to mount and its performance can be significantly increased with minimal adjustments from the attacker. Generally, the jamming strategy adopted in the wild for civilian devices is rather straightforward: a high-power transmitter with a repeating frequency pattern and modulation content sweeps across the relevant \gls{gnss} frequency bands. This is largely due to the low complexity of the interference transmitters and their low-cost. Practically, the majority of the commonly available jammers are \gls{cw} or pulsed jammers, with few exceptions when a simple BPSK PRN modulation is adopted. 

% Civilian jammers produce signals that are simple to detect based on the power advantage. A power spectrum analysis will highlight the presence of any transmitter in the band of interest. Nevertheless, the localization based on \gls{rssi} is often cumbersome as it requires calibrating the receiver signal path. Additionally, range estimation based on \gls{rssi} is effective if the transmitter is static, and its output power does not change. This is because the receiver cannot distinguish power variations from changes in relative position between the receiver and the transmitter. In our case, we 

It is important here to observe the following: while the discussion focuses on jamming transmitters, this is a specific case of generic \gls{gnss}-oriented interference transmitters. The attacker could be operating a spoofer or a meaconer, in addition or substitution to the jammer. In such cases, the \gls{gnss} receiver would be able to acquire and track valid satellite signals even with the antenna not facing the sky. Practically, from a power detection point of view, the only difference is that generally spoofers operate at a much lower power, making them more subtle. Nevertheless, the transmitter power is still larger than the legitimate signal one, allowing detection.

\section{Methodology}
\label{section:methodology}

\textcolor{black}{We have three objectives: (a) detection of the jammer, (b) localization of the transmitter, and (c) characterization of the jammer spectral components, using spectral correlation techniques to classify the type of jammer and its behavior, to estimate its influence on a \gls{gnss} receiver in the affected area. This is achieved in three separate steps: (a) coarse detection of the jamming signal based on relative power estimation outlined in \cref{section:jammer_detection} and periodic power components based on the analysis in \cref{section:jammer_identification}, (b) localization of the jammer by horizon scanning as in \cref{section:jammer_localization} and ultimately (c) identification of the specific type based on periodic statistical property analysis as in \cref{section:jammer_identification}. Notably, the three steps can happen in different sequences, but logically, it should proceed by determining the existence of a jammer, localizing it, and determining its type. Nevertheless, determining the jammer presence and type can be performed at the same time.}

\subsection{Jammer detection}
\label{section:jammer_detection}

After filtering, down-conversion, and sampling the jamming and \gls{gnss} signal is modeled as \cref{eq:signal_model_full}. For each sample index $n$ at a sampling frequency of $\frac{1}{T_s}$ the total received signal is the superposition of two components: the legitimate GNSS signal and the jammer one.
% In \cref{eq:signal_model_full}, $X_{\gls{gnss}}[n]$ defines the complex baseband \gls{gnss} signal of amplitude $J_{\gls{gnss}}$. 
\textcolor{black}{The term $A_{\gls{gnss}}$ is the amplitude of the GNSS signal that can be considered constant in the observation interval. $X_{\gls{gnss}}[n]$ is the complex baseband of the generic GNSS signal in the channel of observation. The GNSS signal does not contribute any power to the detector as it is buried in the noise; the total baseband signal, $X(n)$, is the superposition of the GNSS and jammer signal. 
$A_{J}[n]$ is the instantaneous amplitude of the jammer signal and $f[n]$ is the instantaneous frequency of the jammer, both time-dependent. The $f[n]$ component is instantaneously narrowband but can vary largely between observations. Similarly, $A_{J}[n]$ can be time-varying, defining the strength of the jammer in an arbitrary manner and often tailored to attacker intent. 
% While often it is not as rapidly changing as the frequency component, amplitude modulation of the jammer is also possible and for this reason, it is considered as a time-varying component. 
For example, the adversary can rely on power ramping, i.e., gradually increasing its transmission power to reach the point where it denies GNSS reception. } 
Finally, $\psi$ is a random phase offset component and $\eta$ is a bi-variate symmetric Gaussian noise component with $E_{xy}(\eta) = 0$ and standard deviation $\sigma^2$.

\begin{equation}
    X[n] = A_{\gls{gnss}}X_{\gls{gnss}}[n]nT_s + A_{J}[n]e^{j2\pi f[n]nT_s + j\psi} + \eta[n]
    \label{eq:signal_model_full}
\end{equation}

Intuitively, the countermeasure aims at testing if the jammer signal overpowers the \gls{gnss} signal, practically this translates into the following hypothesis:

\begin{equation}
    \mathcal{H} =
    \begin{cases}
        \mathcal{H}_0\ \mathrm{if}\ SNR_{\gls{gnss}} > SNR_{Jammer} \\
        \mathcal{H}_1\ \mathrm{otherwise}
    \end{cases}
    \label{eq:hypothesis-rt}
\end{equation}

On the other hand, the \gls{gnss} signal is always buried in the thermal noise. 
\textcolor{black}{The in-band interference level, for $N$ samples, is defined based on the energy received at the frontend, as in \cref{eq:psd}.}
\begin{equation}
    E_{X} = \frac{1}{N} \sum_{n=0}^{N} |X[n]|^2
    \label{eq:psd}
\end{equation}
Due to the protected nature of the \gls{gnss} spectral allocation in a benign scenario, the power received should be comparable to the thermal noise in the channel. Otherwise, it is safe to assume that the only contribution in measurable received power is due to the jammer, simplifying the hypothesis to detect the presence of the jammer or not as follows:
\begin{equation}
    \mathcal{H} =
    \begin{cases}
        \mathcal{H}_0\ \mathrm{if}\ A_{J} = 0 \\
        \mathcal{H}_1\ \mathrm{if}\ A_{J} \neq 0
    \end{cases}
    \label{eq:hypothesis-rt-actual}
\end{equation}

If more energy is detected, based on \cref{eq:psd}, this means that a transmitter is active in the band, corresponding to $\mathcal{H}_1$ in \cref{eq:hypothesis-rt-actual}.
\textcolor{black}{
For this reason, we adapt \cref{eq:signal_model_full} to only consider the jamming contribution, resulting in \cref{eq:signal_model} which serves as the signal model in the rest of the work. 
}
\begin{equation}
    X[n] = A_{J}[n]e^{j2\pi f[n]nT_s + j\psi} + \eta[n]
    \label{eq:signal_model}
\end{equation}

\subsection{Localization of interference}
\label{section:jammer_localization}
Localization of an interference source in space is unfeasible with a single antenna or without multiple cooperative agents, but this issue can be resolved by repeating the same measurement in different positions around the area of interest. Intuitively the process is as follows. First, the plane reaches a starting position and starts a hovering phase. \textcolor{black}{Our system calculates power spectral density snapshots in all channels at all heading angles and information contained in the signal energy estimation is relevant for a specific direction the antenna is pointed in. The scanning antenna lobe is pointed towards the horizon and rotated incrementally at different headings. Once a scan is completed at one location, the plane transitions to a fixed-wing flight and proceeds to the next scanning location. Given that the surrounding environment is reasonably static (specifically, the position of the interference transmitter is not changing during the scanning period) several scans from different, even if sparse, locations allow obtaining a 2D reconstruction of the RF environment and geo-reference of the transmitter source.} 

\textcolor{black}{Practically, given the directionality of the sampling antenna, two scans at different positions around the interference source are sufficient to locate it. Nevertheless, it is not necessarily possible to conveniently place the scanning positions to make sure the measurements are acquired with sufficient geometrical diversity, as shown in \cite{SpangheroMGPP:C:2024}, showing the limitations imposed by unfavorable scanning locations.  
In principle, more survey points allow a better understanding of the surrounding RF environment (the source of interference/jamming), as they improve the geometric diversity of the survey points themselves. The exact choice of the location of the scanning positions depends on the objective of the specific mission. Experiments in \cite{SpangheroMGPP:C:2024} show that additional survey points increase the mapping accuracy (at the cost of increased power consumption of the VTOL thus reducing flight time and the coverage of the survey; hence requiring a balance between the two objectives).}

The scans from each location are then superimposed based on their location and non-coherent integration of the estimated power densities provides accurate estimates of the transmitter's position on the ground. This results in a global map of the surveyed area highlighting the eventual transmitters. The referencing, stacking, and fusion algorithm is implemented in \cref{algo:mapfusion}.

\textcolor{black}{
The total heatmap $HM_{fuse}$ is obtained by combining the scans performed at each scanning position $P$, where the directional antenna pose is defined by its position $[x,y]$ and heading in space in the local frame of each scanning point $i\in[1,N]$. For each available scan, the received spectrum is normalized based on the model of the reception antenna pattern (as shown in \cite{SpangheroMGPP:C:2024}) and corrected to make sure that for $\psi \in [-\pi, \pi]$ the radiation pattern is symmetric and defined. This synthetic radiation pattern (SRP) is based on interpolation of the radiation pattern provided by the antenna manufacturer and masks the received power to the main axis of the antenna, defined as orthogonal and exiting the ground plane. 
The individual scans are then referenced in the global frame of the mission and accumulated. The number of scans normalizes this integrated view of the scanning area providing the final representation of the surveyed area in $HM_{fuse}$.}

\begin{algorithm}
\caption{Fusion of multiple scans at different headings and locations, as in \cite{SpangheroMGPP:C:2024}.}
\label{algo:mapfusion}

\textbf{Input:} \\
$HM$: cumulative map view \\
$P = [P_1,...,P_N]$ Scanning poses \\ 
$SRP$:  Normalized antenna radiation pattern\\
\textbf{Output:} \\
$HM_{fuse}$ : Final representation of the surveyed area 
\begin{algorithmic}
    \For {$i=0$ \textbf{to}} $N$
        \For {$[x,y]$ in $P$}
            \State $\Delta\psi \gets \psi^\prime_i - \arctan(x-x^\prime_1, y-y^\prime_1)$
            \State $HM(x,y) \gets HM(x,y) + SRP(\Delta\psi)$
        \EndFor
        \State $HM_{fuse} \gets HM/N$
    \EndFor

\end{algorithmic}
\end{algorithm}

\textcolor{black}{The localization of the adversarial transmitter is performed by intersecting the antenna radiation patterns for those headings where the total received energy as defined in \cref{eq:psd}, is non-zero, so that it verifies $\mathcal{H}_1$ hypothesis. The intersection of these areas leads to identifying the area of maximum probability of finding the transmitter, corresponding to the maximums in $HM_{fuse}$. }

\subsection{Identification based on cyclostationarity}
\label{section:jammer_identification}

Most interference transmitters are reasonably well-behaved. Low-end jammers are often swept transmitters or continuous wave modulated. Generally, commercial jammers broadcast signals with a periodic frequency pattern. This intrinsically means that the spectral components of \cref{eq:signal_model}, are periodic and their correlation is also periodic.
Such signals are often referred to as cyclostationary, as their statistical properties are periodic. This can be leveraged to measure the transmitter's properties and classify its nature (e.g., type of jamming signal). The \gls{scd} is often used to perform blind estimation of signals whose cyclostationary components (e.g., the periodic components) are unknown, making it an excellent tool for analyzing unknown jamming signals. Intuitively, the \gls{scd} analyzes periodicity in the power spectral density of an unknown signal by applying a series of \gls{fft}. 

Several methods exist from the literature to perform cyclostationary analysis (\cite{Borio2016,csp-guide-fam,10.11003/JPNT.2018.7.3.147}). We employ here a \gls{scd} calculation based on \gls{fam} as in  \cite{WOS:000436076600033} and \cite{csp-implementation-fam}. 
The \gls{fam} calculation is performed on each block of samples available from the front end and is then snapshot-based. While the same approach can operate on continuous data streams, fast signal variations between sample blocks might not be as evident in a snapshot-based analysis.  
The starting point is a block of complex baseband samples $X$ of generic length $N$. Given a window of length $N'$ and overlap between sub-blocks of $N'-L$ samples (under the constraint that $L<N'$), the first step is to organize the IQ samples of the block X in a matrix $\bar{X}$ where each line is a frame of length $N'$ and $P$ rows, where P depends on the total amount of samples available.

\begin{equation}
    \bar{X} = 
        \begin{bmatrix}
            X[0:N'] \\
            X[L:L+N'] \\
            ... \\
            X[(P-1)L:(P-1)L+N'] \\
        \end{bmatrix}
    \label{eq:matrix_1}
\end{equation}

\textcolor{black}{
A data tapering and shaping window is applied to the data in \cref{eq:matrix_1}, (e.g., a Hamming window of length $N'$ is multiplied element-wise in each row) and FFTs are calculated per line. After the FFT, each row needs to be phase-compensated by the sample delay $L$. This is done in \cref{eq:matrix_FFT} by multiplying (element-wise, indicated by $\otimes$) the FFT result with a pure phase shift of $\omega$, where $\omega$ are the FFT frequencies $[-\pi,...,\pi-2\pi/N']$.
}
\begin{equation}
    W = 
        \begin{bmatrix}
            FFT(\bar{X}[0,:]) \\
            FFT(\bar{X}[1,:]) \\
            ... \\
            FFT(\bar{X}[P,:]) \\
        \end{bmatrix}
    \label{eq:matrix_FFT}
    \rightarrow 
        W' = 
        \begin{bmatrix}
            W[1,:] \\
            e^{j\omega2L} \otimes W[2,:] \\
            ... \\
            e^{j\omega PL} \otimes W[P,:] \\
        \end{bmatrix}
\end{equation}

\textcolor{black}{
We define the vector $s[i,:,:]$, as the result of multiplying $W'$ with its transposed complex conjugate. To make the computation we directly apply \cref{eq:fast-integration}, which gives the full 2-D \gls{scd}, rotated by \SI{45}{\degree}. 
}
\begin{equation}
    S_x = \frac{1}{P}\sum^{P-1}_{i=0} s[i,:,:]
    \label{eq:fast-integration}
\end{equation}

\textcolor{black}{From the \gls{scd} it is possible to detect the presence of the jammer and its type. One remark is important here. Detection is possible based solely on the energy content of the signal at the band of interest as shown in \cite{SpangheroMGPP:C:2024}, and this is very power efficient. Nevertheless, if identification of the type of jammer is to be performed as well, it is possible to join the coarse detection step with the \gls{scd} by detecting the jammer presence estimating the total magnitude of the clyclostationary frequencies, as shown in the following paragraphs.}

We implement the detector from \cite{Borio2016} as our detection statistics. Detection is performed by integration of the \gls{scd} as in \cref{eq:detection-scf}. If the maximum of the \gls{scd} components integrated along the normalized cyclostationary frequencies defined as $D_j(l)$ in \cref{eq:detection-scf}, is higher than a threshold $T_j$, the jammer is effectively detected.
\begin{equation}
    D_j(l) = [\frac{1}{N'}|\sum^{N'}_{n=0}S_x[n,l]|]_{l \in [0,1]}
    \label{eq:detection-scf}
\end{equation}

\textcolor{black}{Given the AWGN nature of the non-jamming signal, the \gls{scd} integration along the cyclostationary frequencies leads to very small integration values. As the correlation outside the cyclostationary characteristic components is very low, the integration of the spectral correlation of thermal noise is zero-summing, and the processing gain in the cyclostationary components is very high. On the contrary, in the case $\mathcal{H}_1$, the integration will have peaks at the characteristic cyclic frequencies of the jammer. 
} 
The overall algorithm including the peak tracking and detection is shown in \cref{alg:cap}, where the \gls{fam} is calculated first, and for those blocks that exceed the detection threshold the peak tracking algorithm is also applied.

Nevertheless, this method is oblivious to the shape of the cross-correlation or the number of peaks it presents. For example, the cross-correlation for a static continuous wave jamming signal will likely be a single narrow-band peak at the transmitter's frequency (calculated as offset to the center frequency of the receiver). For a \gls{bpsk} jammer, the \gls{scd} will present four peaks at the modulation nodes. This information can be effectively used to detect the type of jamming signal the transmitter is using. 
%There are challenges to this method: based on \cref{eq:detection-scf}, the detector looks for a maximum correlation frequency shift and non-coherently integrates the signal along that shift. 

\textcolor{black}{For this purpose, a better view of the spectral correlation is provided by the \gls{coh}, a normalized version of the \gls{scd}, which is useful to highlight the cyclostationary features of the signal, independently of the signal's power itself \cite{gardner1994cyclostationarity}. A general definition of the \gls{coh} is provided in \cref{eq:coh}, for \gls{scd} $S_x$ of the signal $X$ with length $N$ samples divided into sample blocks of length $N'$, integer divisor of $N$, where $f$ is the sampling channel center frequency and $\alpha$ is the fractional part of the sampling frequency (which defines the resolution of the \gls{scd} computation by defining the minimum resolvable cyclostationary frequency). $C_x^0(f+\alpha/2)$ and $C_x^0(f-\alpha/2)$ are the power spectral density of the signal shifted by $-\alpha/2$ and $\alpha/2$ respectively, where $\alpha$ is one of the fractional parts of the sampling frequency that are used in the \gls{scd} calculation (in our case, the frequencies in the FFT from \cref{eq:matrix_FFT}). 
}

\begin{equation}
    C_x(f,\alpha) = \frac{S_x(f,\alpha)}{\sqrt{C_x^0(f+\alpha/2)C_x^0(f-\alpha/2)}}
    \label{eq:coh}
\end{equation}

Now, if several signals are overlapping, e.g. a \gls{cw} and a \gls{bpsk} modulated signal at the same center frequency the \gls{scd} highlight the individual spectral components of each signal. This simplifies the identification compared to an FFT power/time analysis. There is one relevant remark: signals that change at a rate fast enough to create aliasing in the sampling (e.g. the sweep rate of the jammer is faster than the time required by the front end to collect the sample) will create additional structures in the SCD. Still, the base structure is revealing of the original transmitted signal and its characteristics, as we show in \cref{sec:evaluation}.

\begin{algorithm}
\caption{Peak detection/tracking in snapshot systems}
\label{alg:cap}
    \textbf{Input:} \\
    $N_{FFT}$ : length of the FFT window\\
    $\alpha$ : number of frequency subdivisions in the FAM calculation\\
    $N$ : Number of complex samples per snapshot\\
    $T_h$ : detection threshold for peak acquisition \\
    \textbf{Output:} \\
    $P$ : vector of peaks position as ${|pk|,\Bar{p}}$, magnitude and frequency center of the peak \\
    $S_x$ : spectral correlation of the sample slice \\
    \textbf{Step 1: Memory allocation} 
\begin{algorithmic}
    % \State \textit{\#Interleaved IQ samples, real/complex}
    \State $x \gets BB_{IQ}^{TOW}[0:2*N]$ 
    \State $P \gets [0]$
\end{algorithmic}
\textbf{Step 2: FAM calculation and peak detection} 
\begin{algorithmic}
    \While{True}
        \If{$TOW == \mathrm{new}(TOW)$}
            \State \textit{\#Sample deinterleaving and conversion to complex}
            \State $\Bar{x} \gets BB_{IQ}^{TOW}[::2] + i*BB_{IQ}^{TOW}[1::2]$
            \State \textit{\#FAM calculation}
            \State $S_x \gets \mathrm{FAM}(\Bar{x}, N_{FFT}, \alpha)$ 
            \State \textit{\#Peak extraction}
            \State $PK \gets \mathrm{findPeaks}(S_x)$
            \State \textit{\#Returns [$|p|,F_c$] for each peak}
            \For{$pk$ in $PK$}           
               \If{$|pk| \geq T_h$}
                    \State \textit{\#Jamming detected}
                    \State $\Bar{pk} \gets \mathrm{trackPeak}(pk)$
               \Else
                    \State \textit{\#Jamming not detected}
               \EndIf
            \EndFor
        \EndIf
    \EndWhile
\end{algorithmic}
\end{algorithm}

Additionally, swept jammers present the same \gls{scd} pattern as a \gls{cw} jammer, but it changes over time as the center frequency of the transmitter is shifted. This can be achieved using a peak tracking algorithm. For each frame where the \gls{scd} crosses the detection threshold, a peak finding algorithm marks all the peaks that exist and crosses the threshold. Second, the peak information is stored between samples. Once a new frame is available, the system tracks the new peak by calculating the minimum distance between the previous peaks and the newly detected maximums. If two peaks are close to each other over time, it is reasonable to assume they are the same jammer. Alternatively, a template-matching classification approach was proven to be robust in classifying separate transmitters, but this requires a fitting model and a library of templates \cite{10.11003/JPNT.2018.7.3.147}. 

As a drawback, while the \gls{scd} allows identification of the jammer using parameters like bandwidth of the jamming signal, the shape of the pulses (or modulation), and sweeping bandwidth with the peak tracking approach, it only provides limited information regarding the temporal characteristics of the jammer. As the signal is cyclostationary, as shown in the \gls{scd}, its autocorrelation is also periodic. Hence, we can calculate the following detection statistics for the sample blocks where the detector in \cref{eq:detection-scf}, triggers a jamming detection.

\textcolor{black}{Given the signal used in the \gls{scd} calculation, we define the autocorrelation as in \cref{eq:autocorr}, where the baseband signal $X$ is correlated ($*$ denotes the correlation operator) with a delayed copy of itself at delays $\delta \in \mathcal{D}=[\frac{-N}{2},0[ \cup ]0,\frac{N}{2}]$. If any other peak is present (excluding the correlation with $\delta=0$), it will be located at the jammer cyclic frequencies.}

\begin{equation}
     C_X = X[n]*X[n-\tau]
     \label{eq:autocorr}
\end{equation}

The decision statistic then looks at the peaks in the autocorrelation with the following decision statistic, as defined in \cref{eq:decision-statistics}.
\begin{equation}
    D_c = \max_{\delta \in \mathcal{D}} \frac{1}{N}{\sum^N_{n=0}|X[n]*X[n-\tau]|}
    \label{eq:decision-statistics}
\end{equation}

For swept or single-tone jammers, the decision statistics \cref{eq:decision-statistics}, will present peaks at jammer sweeping periods, as a function of the receiver sampling frequency. While the \gls{scd} and its normalized version \gls{coh} give information on the spectral structure, the signal's autocorrelation completes and augments this information with more insight into the periodic properties. In the case of swept jammers, from \cref{eq:autocorr}, the detector extracts the period of the sweeping, while \cref{eq:fast-integration}, provides simultaneous information on the center frequency, bandwidth of the swept signal, and in case of more complex signal structures, the type of the modulation.

\section{Experimental setting and platform}
\label{sec:experiment_setup}

Experiments involving interference in the protected navigation bands are illegal without the relevant authorities' authorization, and obtaining permission to conduct such tests is complex. 
To overcome these limitations, we used two separate approaches. Calibration measurements are performed first in a shielded environment where we can broadcast simulated and real GNSS constellation signals and jamming waveforms without disturbing nearby devices. 

Measurements with actual \gls{gnss} jammers were performed at Jammertest 2023 in Norway \cite{jammertest-event}, where over-the-air transmission of interference signals was possible. The data collected is post-processed using the approach from \cref{section:methodology} to detect, localize (when possible) and identify the offending transmission source. Different \gls{gnss} jammer devices are tested, in multi-frequency and multi-constellation settings.  

The live experiments were conducted in an open field, where different jammers were distributed. The drone takeoff area and initial flight path are defined so that the initial operations are in a benign scenario, to avoid any material damage to the experimental platform. The flight path is handled by the automatic flight control system, which is dependent on the flight computer and the \gls{gnss} systems, while the selection of the hovering points is done manually by the operator. 

A preliminary test is performed in a controlled environment using the Safran Skydel \gls{gnss} simulator. The simulator frontend is based on Ettus USRP X300 with 2x2 Transmission paths. The signals are combined in a single output and transmitted over a calibrated cable to the recording system. The recorded baseband signals are used as representative samples of the \gls{scd} response to jammer waveform and time/frequency behavior.
We test a battery of common jamming signals at L1 based on the specifications of the real jammers provided in Jammertest \cite{transmission-plan}, as described in \cref{tab:jammer-testings-lab}.

\begin{table}[h!]
\centering
\caption{Jamming modulations, carriers and signal parameters}
\begin{tabularx}{\linewidth} { 
  | >{\raggedright\arraybackslash}r
  | >{\centering\arraybackslash}r 
  | >{\raggedleft\arraybackslash}X 
  | >{\raggedleft\arraybackslash}X |
  }
 \hline
 Type & Center Freq. & Bandwidth & Sweep time\\
 \hline
 \gls{cw}\_1 & \SI{1575.42}{\mega\hertz} & Single tone & - \\ 
 \hline
 Chirp\_N & \SI{1575.42}{\mega\hertz} & \SI{1}{\mega\hertz} & \SI{100}{\micro\second} \\
 \hline 
 Chirp\_W & \SI{1575.42}{\mega\hertz} & \SI{5}{\mega\hertz} & \SI{100}{\micro\second} \\ 
 \hline
 Chirp\_W\_O & \SI{1577.92}{\mega\hertz} & \SI{10}{\mega\hertz} & \SI{100}{\micro\second} \\ 
 \hline\hline
 Type & Center Freq. & PRN & Chip rate \\
 \hline
 \gls{bpsk}\_W & \SI{1575.42}{\mega\hertz}& \SI{8}{\mega\hertz} & \SI{1.023}{\mega Chip/\second} \\

 \hline

\end{tabularx}
\centering
\label{tab:jammer-testings-lab}
\end{table}

During the Jammertest field testing, different types of jammers are deployed in the field with varying power ranges and frequency patterns. The test is split into two parts: a static test to validate the laboratory experiments and a dynamic jammer localization test. 

\begin{figure}
    \includegraphics[width=\linewidth]{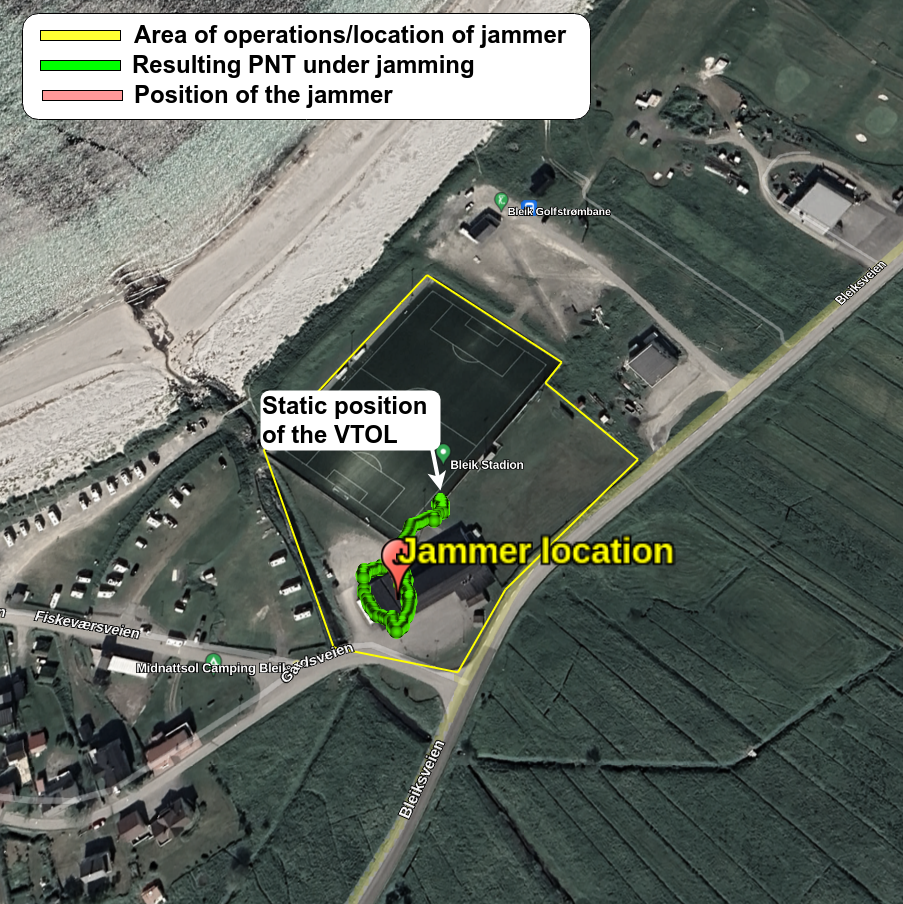}
    \caption{Static test: the initial \gls{pnt} solution degrades with the jammer action. \textcolor{black}{The PNT of the victim receiver, influenced by the jammer, is marked in green. The area of VTOL operations is delimited in yellow.}}
    \label{fig:jammer-static}
\end{figure}

\textit{Static tests in jamming and spoofing - } The configuration of the experiment is shown in \cref{fig:jammer-static}. A transmission antenna is positioned on top of the building as marked in \cref{fig:jammer-static}, and the \gls{vtol} is placed in the nearby field. While these measurements are not useful for the localization of the jammer, as the \gls{vtol} is static, they provide a baseline for the receiver's frontend capabilities and the validity of the method when applied to the receiver-provided IQ samples over the air. The jamming signals are generated with a Safran Skydel setup operated by the Jammertest team and equivalent to the one used in our laboratory analysis. We remark that being in close proximity to the jammer the PNT of the receiver is influenced by jamming, but given the static nature of the test setup this does not constitute a safety concern. 

\begin{figure}
    \includegraphics[width=\linewidth]{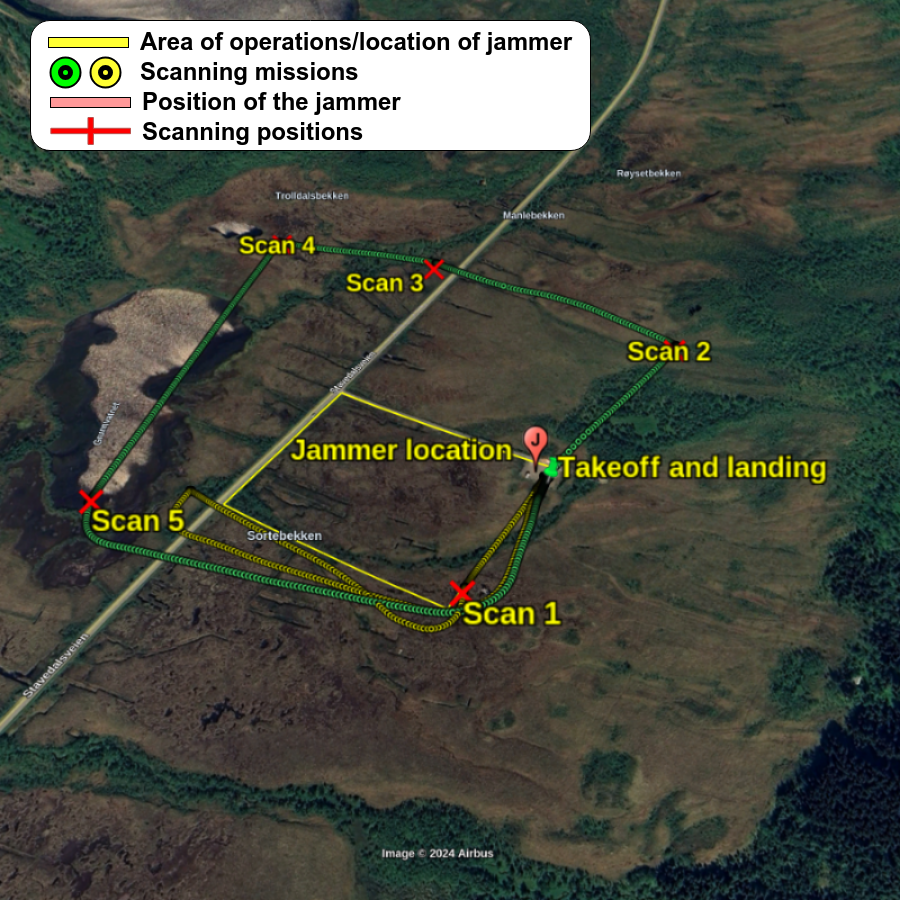}
    \caption{Scanning mission flight path and test area layout. \textcolor{black}{The jammer positioning is within the area of operations marked in yellow (solid). Two missions are performed, a 3-point survey (yellow, dotted) and a multi-point survey (green, dotted).}}
    \label{fig:live-jammer-flight}
\end{figure}

\textit{Flight tests in jamming}: dynamic jamming tests concern and actual flight paths with several scanning transitions. 
\textcolor{black}{during dynamic tests the \gls{vtol} performs an actual flight mission in the proximity of a denied area with several flight-to-scan transitions, around the location of the jammer.}
The tests use commercially available jammers with a wideband chirp or \gls{cw} transmission. The devices selected for the testing are low-cost, highly-effective jammers such as a \gls{ppd} and are a significant problem for the robustness and reliability of navigation systems. Two jammers are located in the area marked in \cref{fig:live-jammer-flight}, with different frequency patterns summarized in \cref{tab:jammer-testings-live-ppd}. The overall test area is about \SI{3}{\kilo\meter^2} with a smaller search area of about \SI{1}{\kilo\meter^2}.  

\begin{table}[h!]
\begin{center}
\caption{Real-life jammers used in the dynamic open-air test scenarios. These devices are commonly available for purchase.}
\begin{tabularx}{\linewidth} { 
  | >{\raggedright\arraybackslash}r
  | >{\centering\arraybackslash}r
  | >{\raggedleft\arraybackslash}r 
  | >{\raggedleft\arraybackslash}X
  | >{\raggedleft\arraybackslash}X |
  }
 \hline
 Type & Center Freq. & Bandwidth & Sweep time & Power \\
 \hline
 LP\_Swept & \SI{1575}{\mega\hertz} GPS L1 & \SI{15}{\mega\hertz} & \SI{6}{\micro\second} & \SI{18}{\dB m}\\ 
 \hline
 LP\_Multi\_1 & \SI{1575}{\mega\hertz} GPS L1 & \SI{20}{\mega\hertz} & \SI{13}{\micro\second} & \SI{30}{\dB m}\\
 \hline 
 LP\_Multi\_2 & \SI{1227}{\mega\hertz} GPS L2 & \SI{14}{\mega\hertz} & \SI{13}{\micro\second} & \SI{30}{\dB m}\\
 \hline 
 LP\_Multi\_3 & \SI{1176}{\mega\hertz} GPS L5 & \SI{17}{\mega\hertz} & \SI{13}{\micro\second} & \SI{30}{\dB m}\\
 \hline

\end{tabularx}
\label{tab:jammer-testings-live-ppd}
\end{center}
\end{table}

\begin{figure}
    \centering
    \includegraphics[width=\linewidth]{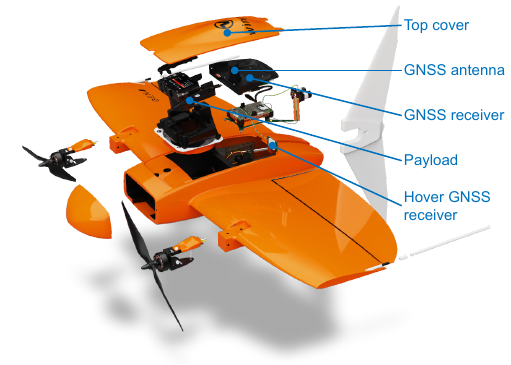}
    \caption{WingtraOne GenII \gls{vtol} used to test the scheme in the field trials \cite{wingtraone}.}
    \label{fig:vtol-plane}
\end{figure}

Our experimental platform is based on the WingtraOne GenII \gls{vtol}, in \cref{fig:vtol-plane}. This drone is used for precision photogrammetry and land surveying over large areas. The \gls{gnss} receiver used for navigation and measurement collection is the Septentrio Mosaic X5 multi-constellation, multi-frequency \gls{gnss} receiver. The \gls{vtol} platform uses a top-mounted antenna that is used, during normal flight mode, for navigation and, in hovering mode, for measurement of the power density in the \gls{gnss} bands. 

The Mosaic X5 receiver precision navigation information during flight mode can be enhanced with Post-Processing-Kinematics to achieve centimeter-level accuracy for the flight path. The \gls{gnss} \gls{pnt} solution rate is set to \SI{10}{\hertz} while the baseband samples are provided with a rate of \SI{0.5}{\hertz}, synchronously in all 3 bands of operation of the Mosaic X5 receiver.
While the IQ samples are provided too sparsely to be used to track the signal, they provide a meaningful snapshot of the frequency content in the \gls{gnss} channel. The antenna used for precise navigation and spectrum sensing is a triple-band helical antenna with an average beam width of \SI{60}{\degree} over the entire L-band spectrum. 

The normal \gls{vtol} operation can be in fully autonomous mode: the user configures an area to be surveyed and generates a flight path. For practical equipment safety reasons, the \gls{vtol} flew in semi-autonomous mode during the interference scanning mission. While the flight controller stabilizes the \gls{vtol} attitude, the operator can manually decide when and where to trigger the transition to hovering mode and start a measurement. In hovering mode, the \gls{vtol} is capable of maintaining a stable position vertically, where the heading of the main plain of the \gls{vtol} can be turned at precise increments. In this mode, the main lobe of the navigation antenna is slightly tilted downwards (approximately \SI{7}{\degree}, due to the \gls{vtol} frame structure) and it is used to detect transmitters on the ground. As the main \gls{gnss} receiver is used for interference scanning during measurement mode, a secondary \gls{gnss} receiver and inertial navigation system measurements (INS) mounted within the flight controller are used to obtain the location of the measurement point and orientation of the measurement antenna.

\section{Evaluation and results}
\label{sec:evaluation}

\begin{table*}[h!]
\centering
\caption{Summary of the performed tests in different settings}

\begin{tabularx}{\textwidth} { 
  | >{\raggedright\arraybackslash}X
  | >{\raggedleft\arraybackslash}p{0.5in} 
  | >{\raggedleft\arraybackslash}p{0.5in} 
  | >{\raggedleft\arraybackslash}p{1in} 
  | >{\raggedleft\arraybackslash}p{0.5in} 
  | >{\raggedleft\arraybackslash}p{0.5in} 
  | >{\raggedleft\arraybackslash}p{0.55in} 
  | >{\raggedleft\arraybackslash}p{0.55in} 
  | >{\raggedleft\arraybackslash}X |
  }
 \hline
 Test & Center Freq. & Jammer & Setting & Type & Detect and Identify & Localization & Platform & Reference \\
 \hline
 Lab\_test\_1 & GPS L1 & Chirp\_W\_O & Shielded chamber & Simulated & YES/YES & N/A & \gls{sdr} & \cref{fig:test-eval-lab-cw,fig:test-eval-lab-cw-coh}\\ 
 \hline
 Lab\_test\_2 & GPS L1 & BPSK\_W & Shielded chamber & Simulated & YES/YES & N/A & \gls{sdr} & \cref{fig:test-eval-lab-prn,fig:test-eval-lab-prn-coh}\\
 \hline 
 OTA\_Static\_1 & GPS L1 & PRN & Over the air & Static & YES/YES & N/A & \gls{vtol} + Mosaic X5 & \cref{fig:ground-jammer-1,fig:ground-jammer-1-coh,fig:degradation-static}\\
 \hline 
 OTA\_Static\_2 & GPS L1 + offset & CW\_tone & Over the air & Static & YES/YES & N/A & \gls{vtol} + Mosaic X5 & \cref{fig:ground-jammer-2,fig:ground-jammer-2-coh,fig:degradation-static} \\
 \hline
 OTA\_Flight\_1 & GPS L1/L2/L5 & LP\_Swept, LP\_Multi & Over the air (localization) & Dynamic & N/A & YES & \gls{vtol} + Mosaic X5 & \cref{fig:live-jammer-convergence,fig:live-jammer-mapfusion}\\
 \hline
 OTA\_Flight\_1\_a & GPS L1/L2/L5 & LP\_Swept, LP\_Multi & Over the air (antenna away from jammer) & Dynamic & NO/NO & YES & \gls{vtol} + Mosaic X5 & \cref{fig:flight-benign-1,fig:flight-benign-1-coh}\\
 \hline
 OTA\_Flight\_1\_b & GPS L1/L2/L5 & LP\_Swept, LP\_Multi & Over the air (antenna towards jammer) & Dynamic & YES/YES & YES & \gls{vtol} + Mosaic X5 & \cref{fig:flight-jammer-weak,fig:flight-jammer-weak-coh,fig:ota-dyn-hard-eval}\\
 \hline

\end{tabularx}

\label{tab:summary-of-test}
\end{table*}

The laboratory simulation-based evaluation is not as comprehensive as the live-sky testing, but the analysis is provided here as a reference and comparison point for the method when applied to the samples provided by the \gls{gnss} front end. Additionally, it shows that, although \gls{sdr}-based sampling provides better results in terms of the noise floor and streaming rate, the amount of data to be processed is overwhelming for low-end, mobile platforms. Conversely, the snapshots provided by a capable \gls{gnss} receiver already contain sufficient information for our method to provide meaningful detection and situational awareness at a rate that is easily manageable even in low power embedded platforms. For convenience, a summary of the tests performed in different settings is provided in \cref{tab:summary-of-test}, where the different jammers follow the same naming as in \cref{tab:jammer-testings-lab,tab:jammer-testings-live-ppd}. \textcolor{black}{The results for detection, localization, and identification are presented in this order, but there is no limitation to the combination of the operations, that can be executed out of order. For example, the device could perform identification right after detection and before the entire survey is completed.}

\textit{Preliminary evaluation}: The jammers in \cref{tab:jammer-testings-lab} are simulated and sampled with our \gls{sdr}-recorder. Specifically, the cases of a chirp and a \gls{bpsk} PRN-modulated jammer are interesting as experimental evidence shows they have the best chances of effectively jamming the receiver. Additionally, the Skydel simulation setup reflected the real positioning of the receiver and jammer in the field test. The jammer successfully denied the \gls{pnt} solution at the receiver, and even for the attacks where the receiver still provides a \gls{pnt}, the solution is degraded enough to be unsuitable for navigation purposes. The evaluation is performed with a single jammer at the L1 carrier (or at a close-by offset), but multiple jammers can be deployed and recognized in the \gls{scd} representation.

\begin{figure*}
    \centering
    \begin{tabular}{@{}p{0.25\textwidth}@{}p{0.25\textwidth}@{}p{0.25\textwidth}@{}p{0.25\textwidth}@{}}
    \subfloat[]{
        \includegraphics[width=\linewidth]{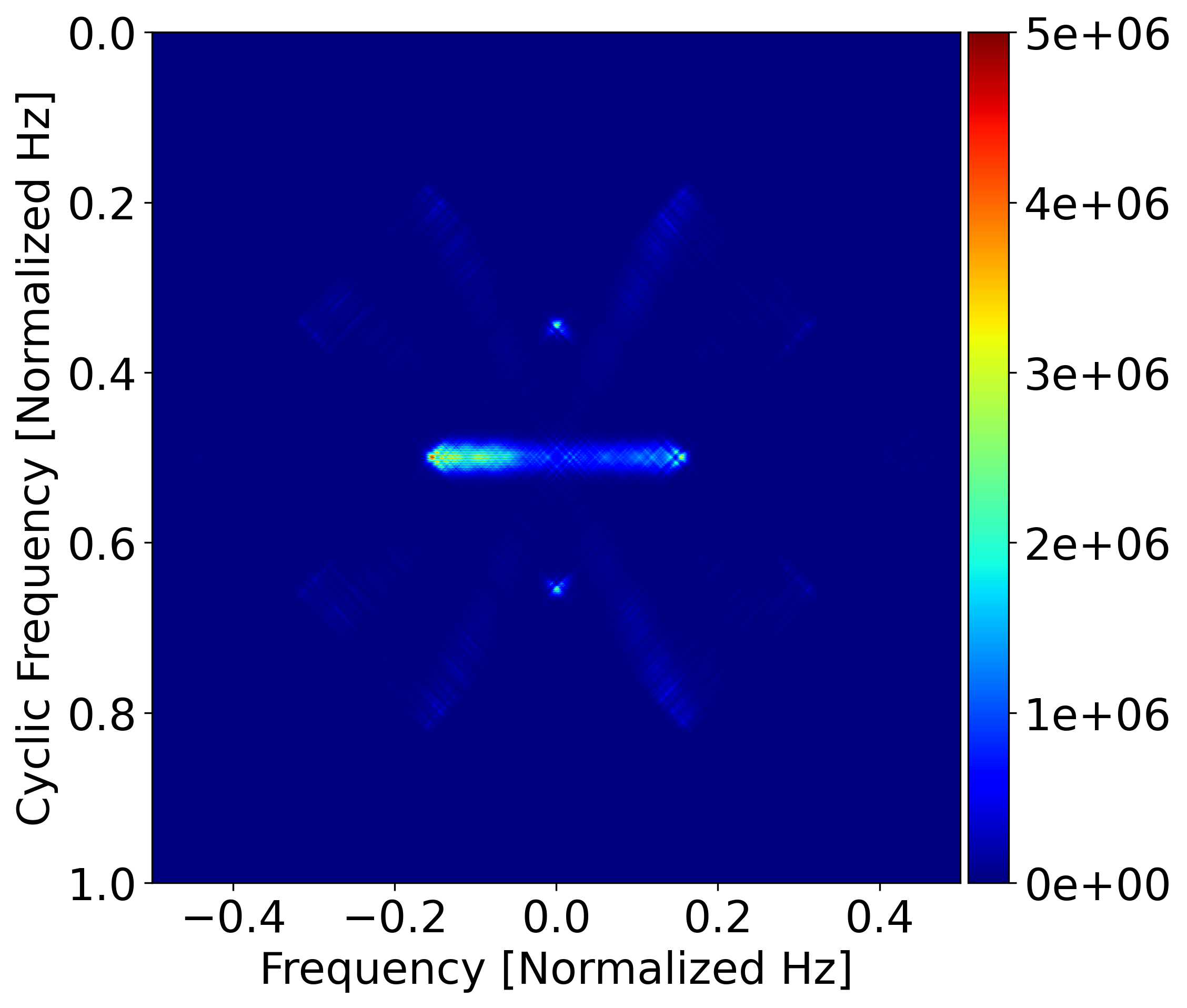}
        \label{fig:test-eval-lab-cw}
    }&
    \subfloat[]{
        \includegraphics[width=\linewidth]{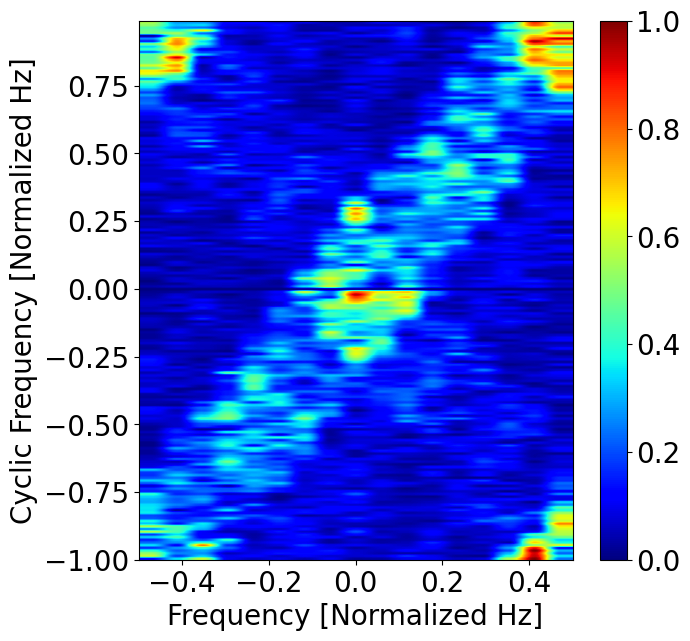}
        \label{fig:test-eval-lab-cw-coh}
    }&
    \subfloat[]{
        \includegraphics[width=\linewidth]{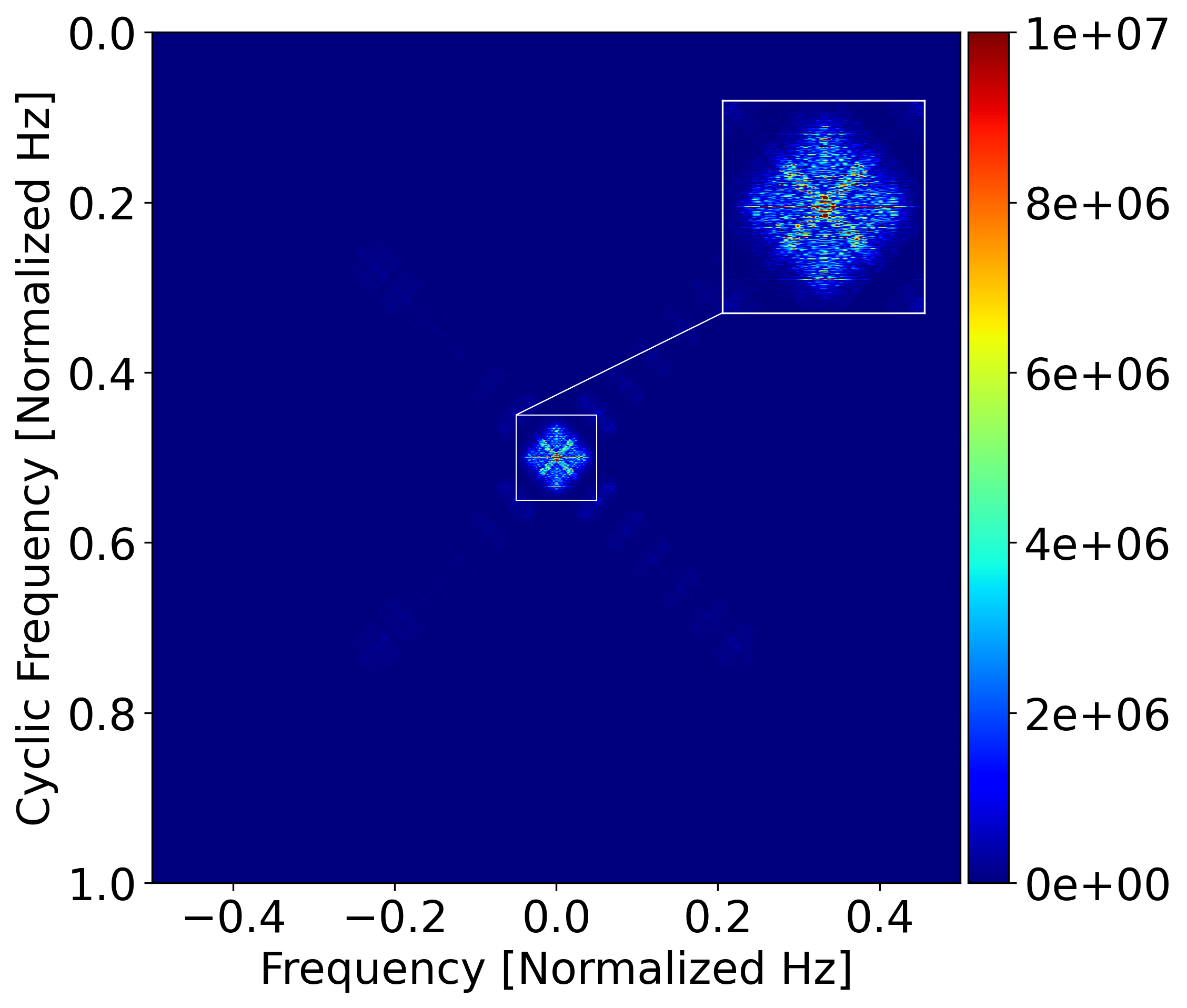}
        \label{fig:test-eval-lab-prn}
    }&
    \subfloat[]{
        \includegraphics[width=\linewidth]{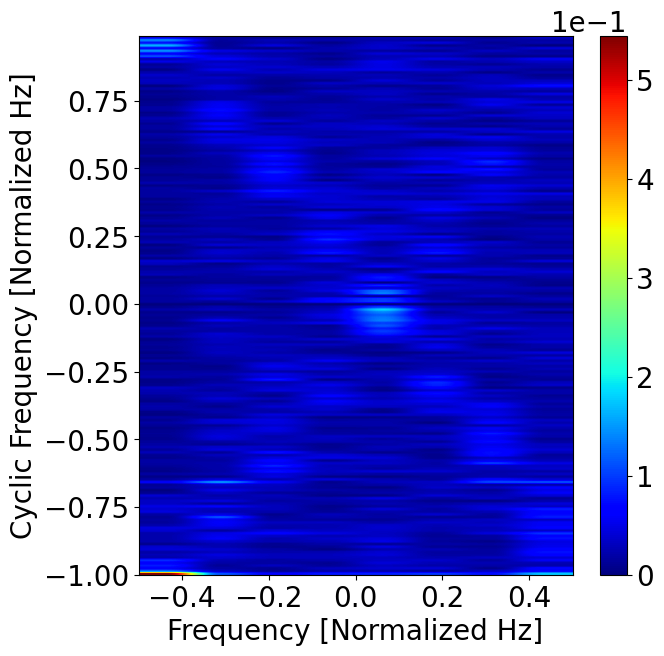}
        \label{fig:test-eval-lab-prn-coh}
    }%
    \end{tabular}
    \caption{Lab\_test: \gls{sdr} based sampling of representative examples from \cref{tab:jammer-testings-lab}, specifically Chirp\_W and \gls{bpsk}\_W. Lab\_test\_1: \gls{fam} of a swept jammer (a) and \gls{coh} structure (b); Lab\_test\_2: \gls{fam} of \gls{bpsk} PRN 11 modulated jammer (c) and \gls{coh} structure of the \gls{bpsk} jammer (d).}
    \label{fig:lab-eval-jammers}
\end{figure*}

In \cref{fig:test-eval-lab-cw,fig:test-eval-lab-prn} the \gls{scd} plots of the two experiments are shown, with the respective coherence functions in \cref{fig:test-eval-lab-cw-coh,fig:test-eval-lab-prn-coh}. The figure scale is a linear adimensional representation of the relative power level and not a received signal strength - because there is no gain calibration at the front end or antenna and the received IQ samples are not normalized by the front-end gain (this is valid in all \gls{scd} representations in the text, unless otherwise specified). Clear structures are present in \cref{fig:test-eval-lab-cw}, where a chirp tone is used to jam the center of the L1 carrier. It can be seen in \cref{fig:test-eval-lab-cw} how the \gls{scd} resolution is limited by the sampling rate. The tested jammer is the Chirp\_N jammer from \cref{tab:jammer-testings-lab}. To achieve comparable results to the Mosaic \gls{gnss} receiver, we limit the sample block length ($N$ in \cref{eq:matrix_FFT}) to 2048 samples, which corresponds to a temporal resolution between frames of \SI{133}{\micro\second} at \SI{15}{\mega S/\second} sampling rate. This causes artifacts in the SCD due to the jammer chirp tone aliasing in the measurement. Practically, the effect is due to the fast movement of the peak in the spectrum. Although a simple solution would be to increase the sampling frequency, this causes a significant increase in computational power. Nevertheless, it is still possible to track the peaks of the jammer even with aliasing artifacts, as the \gls{coh} value of the real peak is higher than the aliasing one, as shown in \cref{fig:test-eval-lab-cw-coh}. 

\textcolor{black}{Aliasing effects} are not visible if the jammer is static at a certain frequency such as the one in \cref{fig:test-eval-lab-prn} in the case of a PRN-modulated L1 jammer. In this case, the \gls{scd} representation is even more revealing. The characteristic four peaks of the \gls{bpsk} modulation is recognizable from \cref{fig:test-eval-lab-prn}, and the spectral structure in \cref{fig:test-eval-lab-prn-coh}. The strength of the jammer is well represented in the \gls{fam} in \cref{fig:test-eval-lab-cw,fig:test-eval-lab-prn}, where higher relative values show the presence of a stronger interference source. From the detection perspective, the hypothesis test in \cref{eq:hypothesis-rt-actual}, is straightforward as the integration of the \gls{fam} over the normalized cyclic frequencies directly shows if a jammer is present, as shown in \cref{eq:detection-scf}. 

% \begin{figure*}
%     \centering
%     \begin{subfigure}[t]{.255\textwidth}
%         \includegraphics[width=\linewidth]{IQ_1_0_FAM_sweep.png}
%         \caption{Lab\_test\_1: \gls{fam} of a swept jammer}
%         \label{fig:test-eval-lab-cw}
%     \end{subfigure}%
%     \begin{subfigure}[t]{.255\textwidth}
%         \includegraphics[width=\linewidth]{IQ_1_0_COH_sweep_rev.png}
%         \caption{Lab\_test\_1: \gls{coh} structure in swept jammer}
%         \label{fig:test-eval-lab-cw-coh}
%     \end{subfigure}%
%     \begin{subfigure}[t]{.25\textwidth}
%         \includegraphics[width=\linewidth]{IQ_1_0_FAM_BPSK.png}
%         \caption{Lab\_test\_2: \gls{fam} of \gls{bpsk} PRN 11 modulated jammer}
%         \label{fig:test-eval-lab-prn}
%     \end{subfigure}%
%     \begin{subfigure}[t]{.245\textwidth}
%         \includegraphics[width=\linewidth]{IQ_1_0_COH_chirp_rev.png}
%         \caption{Laboratory test: \gls{coh} structure of a \gls{bpsk} jammer}
%         \label{fig:test-eval-lab-prn-coh}
%     \end{subfigure}%
%     \caption{Lab\_test\_2: \gls{sdr} based sampling of representative examples from \cref{tab:jammer-testings-lab}, specifically Chirp\_W and \gls{bpsk}\_W}
%     \label{fig:lab-eval-jammers}
% \end{figure*}

Similar results are obtained from live testing with \gls{ota} jammers. The results are consistent with the assessment done in the laboratory and the IQ samples provided by the \gls{vtol} GNSS receiver allow rapid detection and jammer-type identification. The distinct patterns and spectral structures in the \gls{scd} are present even with lower-resolution frequency bins. The bandwidth, frequency center, and sweep times are measurable in both the chirp jammer and the PRN jammer and are consistent with the ground truth provided during Jammertest. 

\textit{Fixed location jamming detection and modulation analysis}: Ground-based tests at the location in \cref{fig:jammer-static} confirm the observations from the experimental simulation-based tests. Detection of the jammer is effective in all cases and clear also from the degradation of the \gls{pnt} solution, where the initial correct \gls{pnt} solution is progressively lost to the jammer action. 

\begin{figure}
    \centering
    \includegraphics[width=\linewidth]{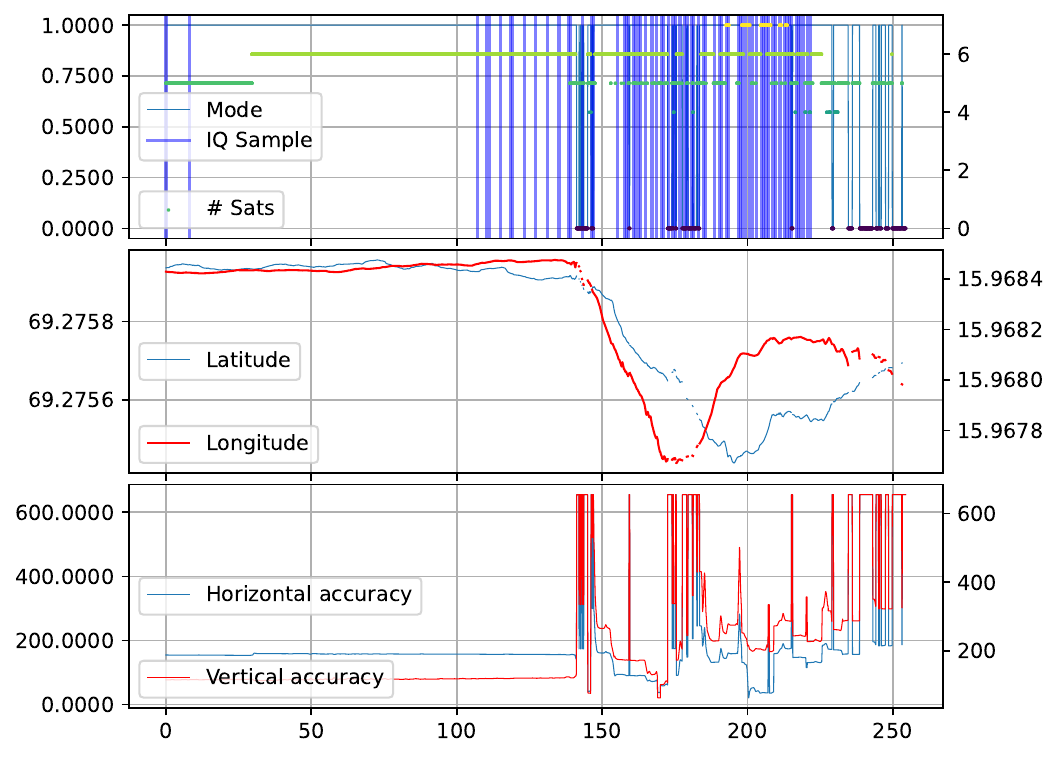}
    \caption{OTA\_Static\_1 and OTA\_Static\_2: degradation and loss of \gls{pnt} due to jammer. The timescale is relative to the beginning of the test.}
    \label{fig:degradation-static}
\end{figure}

\begin{figure*}
    \centering
    \begin{tabular}{@{}p{0.25\textwidth}@{}p{0.25\textwidth}@{}p{0.25\textwidth}@{}p{0.25\textwidth}@{}}
    \subfloat[]{
        \includegraphics[width=\linewidth]{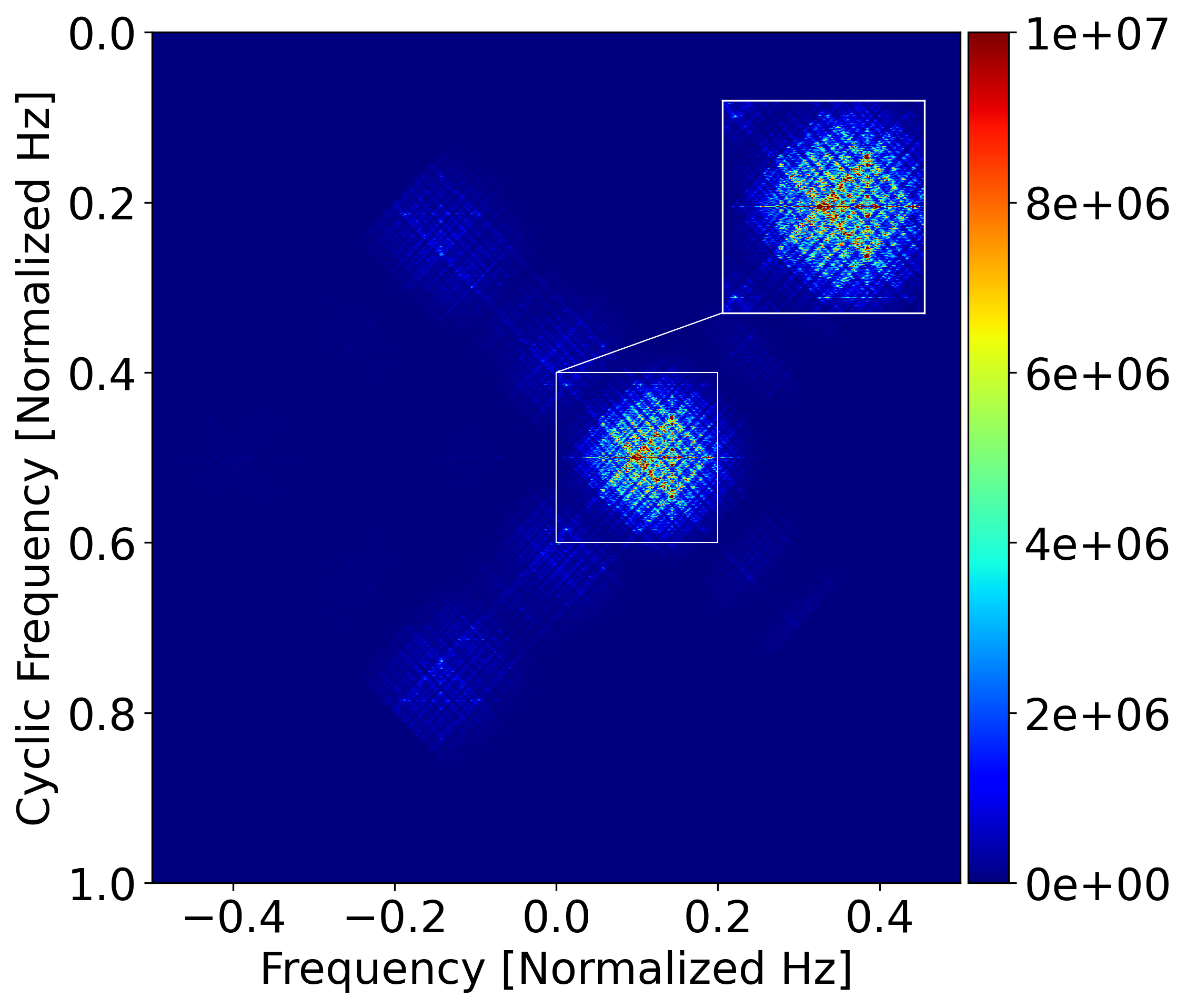}
        \label{fig:ground-jammer-1}
    }&
    \subfloat[]{
        \includegraphics[width=\linewidth]{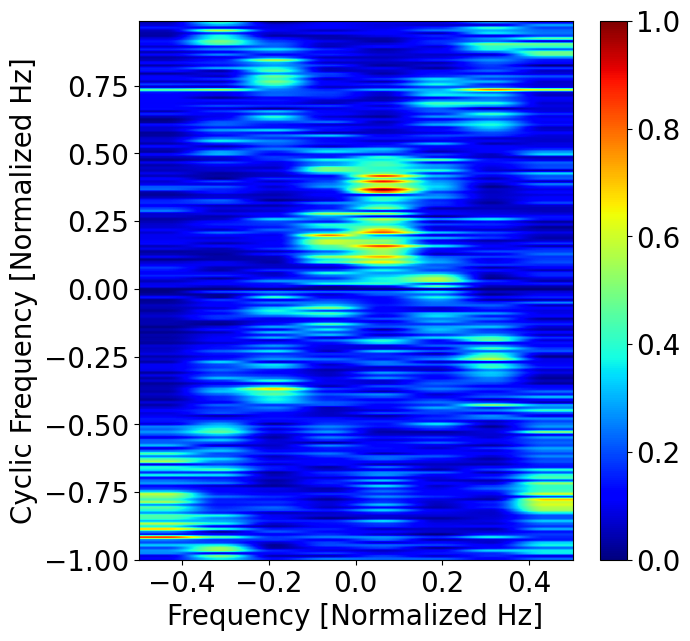}
        \label{fig:ground-jammer-1-coh}
    }&
    \subfloat[]{
        \includegraphics[width=\linewidth]{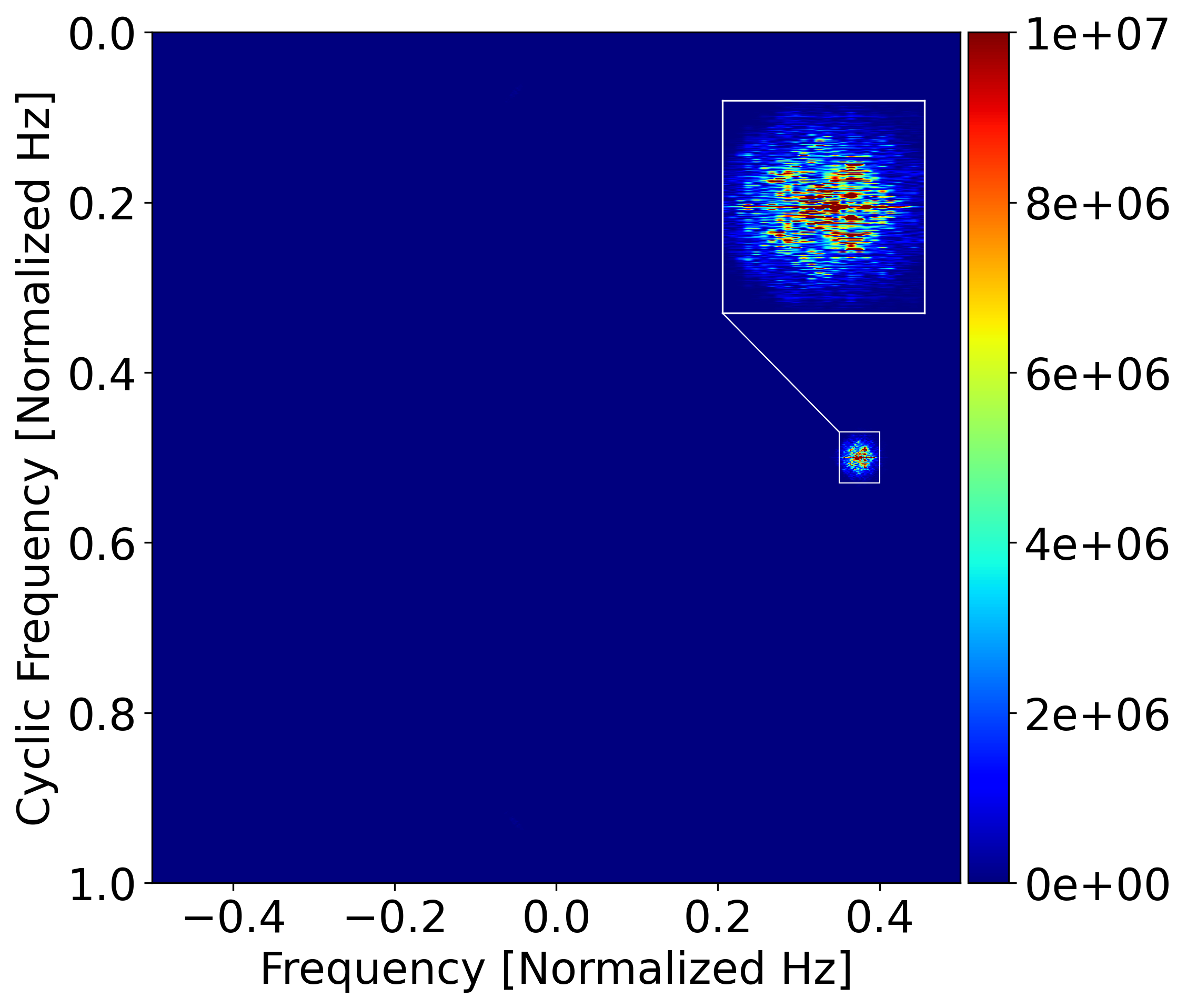}
        \label{fig:ground-jammer-2}
    }&
    \subfloat[]{
        \includegraphics[width=\linewidth]{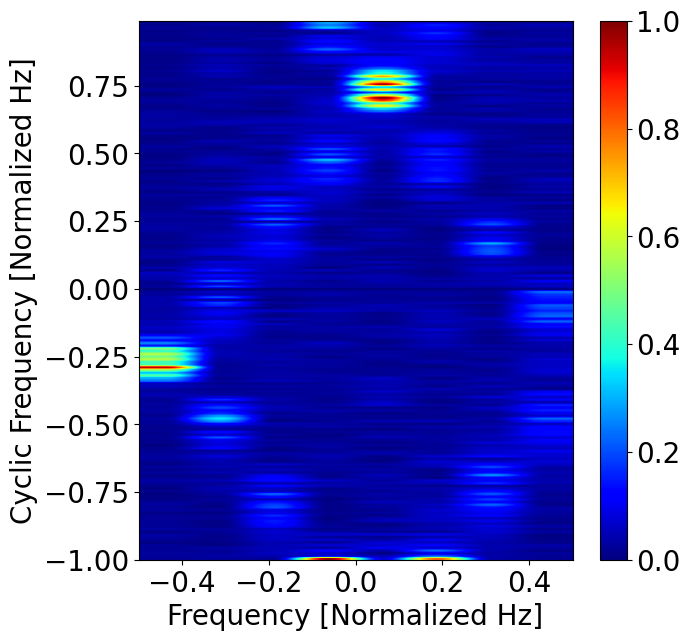}
        \label{fig:ground-jammer-2-coh}
    }
    \end{tabular}
    \caption{OTA jammer with static measurements: ground test evaluation of the Mosaic receiver IQ samples for \gls{scd} generation. The center frequency of the L1 frontend is \SI{1584}{\mega\hertz} for the Mosaic X5 L1 front-end. OTA\_Static\_1: GPS L1 jammer with PRN modulation (a) and L1 jammer spectral coherence (b); OTA\_Static\_2: Tone jammer at an offset with respect to the carrier (c) and tone jammer spectral coherence (d).}
    \label{fig:ota-static-eval}
\end{figure*}

At the onset of the attack, the accuracy of the solution decreases, both in the horizontal and the vertical planes, as can be seen in \cref{fig:degradation-static}, where the position of the victim receiver slowly drifts from the true position as the attack progresses. First, the internal interference mitigation algorithm tries to adapt to the jammer but eventually, the mitigation fails making the receiver's \gls{pnt} solution unreliable and progressively less available. 
During jamming the \gls{gnss} receiver stops providing a valid solution and the attack is successfully detected as in \cref{fig:ground-jammer-1,fig:ground-jammer-2} where two different jammers are detected. This is representative of information content provided by raw IF data from the \gls{gnss} front-end in comparison to \gls{sqm} only measurements. In both cases, even if the receiver cannot produce a \gls{pnt} solution, the platform can still detect, report, and classify the interference, based on the spectral coherence. This is seen in \cref{fig:ota-static-eval}, where the \cref{fig:ground-jammer-1,fig:ground-jammer-2} are two representative samples taken during the periods of activation of the jammer.
Specifically, the PRN jammer in \cref{fig:ground-jammer-1-coh} shows a similar signature as the \cref{fig:test-eval-lab-prn-coh}, but the intensity of the jamming signal during live testing is higher, hence the more structured coherence plot in \cref{fig:ground-jammer-1-coh}. On the other hand, the jamming measured in \cref{fig:ground-jammer-2-coh} is a single-tone jammer at an offset from the L1 carrier and shows a very different signature. These results are consistent with the \gls{sdr}-based ones. In particular, the PRN-modulated jammer shows the same spectral coherence structures, indicating that the Mosaic X5 raw sampler data performs similarly to \gls{sdr}-based sampling.

In the static ground tests, the \gls{vtol} is not actively flying. For this reason, only detection of the jammer is possible. Nevertheless, this is an important result: at the cold start, the system can rely on the IQ information provided by the GNSS receiver to monitor the environment before the \gls{vtol} takes off to avoid operation in a denied environment. If this information is immediately available to the \gls{gnss}-enabled system, the detection and identification of a potential jammer even before a \gls{pnt} solution is available, and afterward provides continuous monitoring of the quality of the RF spectrum.

\textit{In-flight localization and characterization of the interference source}:
The full combination of interference localization and characterization is evaluated in the dynamic tests, where the \gls{vtol} is used for large-scale interference hunting. The flight mission is shown in \cref{fig:live-jammer-flight}, with five separate scans around a location where the jammer is located. Notably, the jammer was disabled at take-off and was enabled only after the \gls{vtol} completed the maneuver and positioned itself further away. This does not limit the validity of the results, but the precaution was taken due to safety concerns for the ground operators and the \gls{vtol} itself. \textcolor{black}{In all dynamic tests, the \gls{vtol} moves around the jammer that stays static within the area marked in \cref{fig:live-jammer-flight}. This is due to the nature of the test area, which requires the jammers to be located at a fixed position.} 

\begin{figure*}
\centering
    \subfloat[]{
        \includegraphics[width=.3\textwidth]{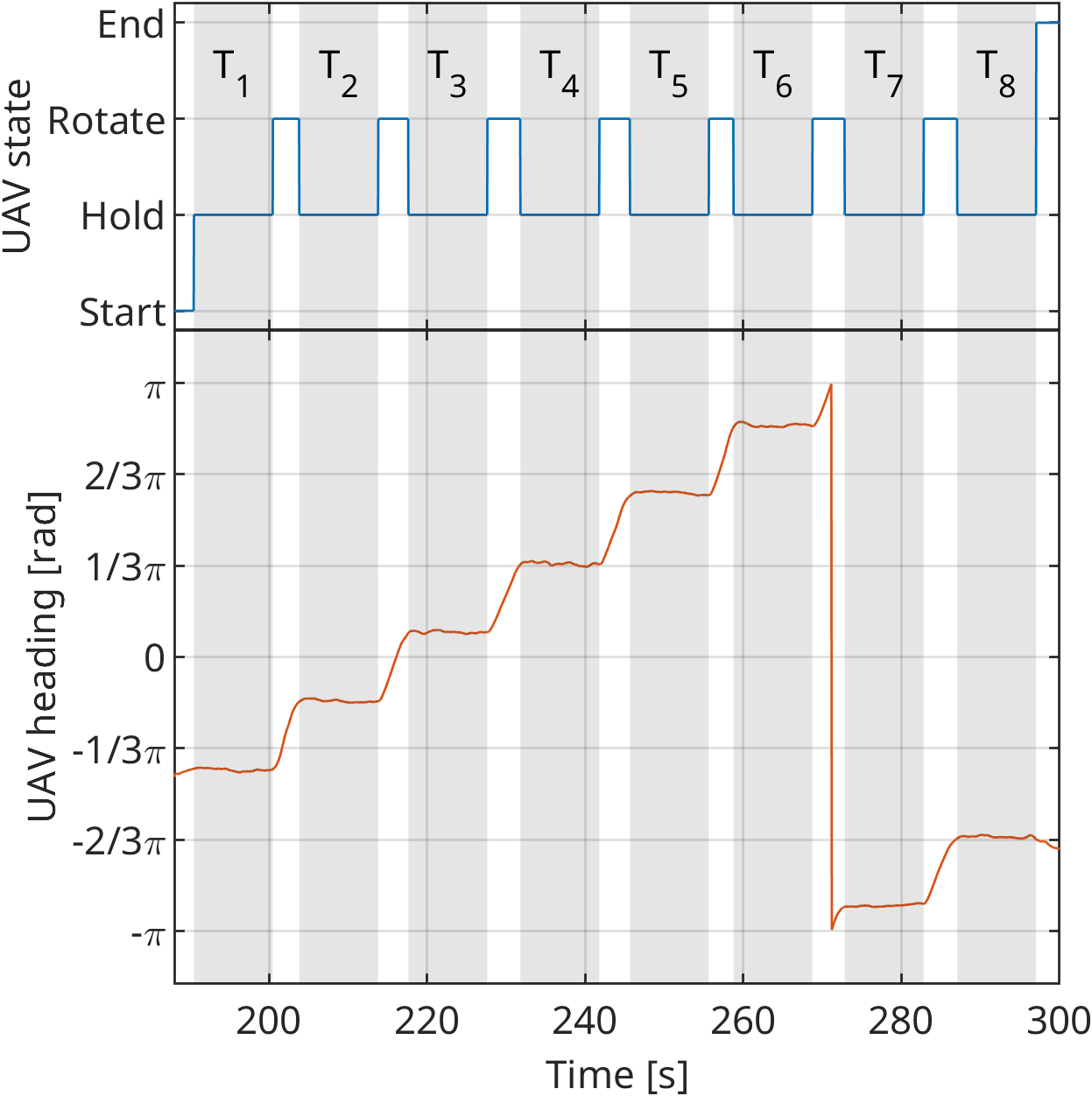}
        \label{fig:flight-scanning}
    }
    \quad\quad
    \subfloat[]{
        \includegraphics[width=.3\textwidth]{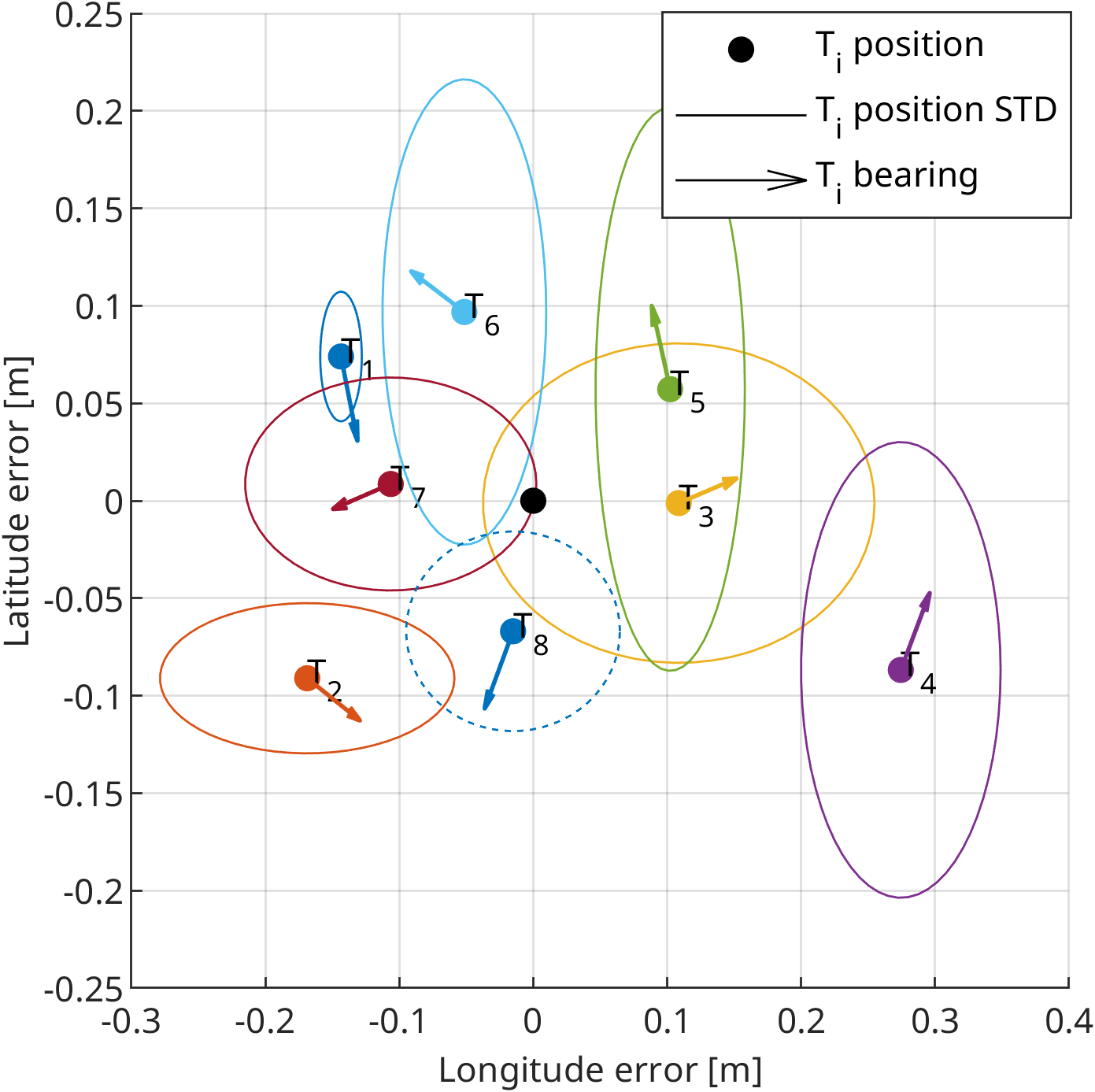}
        \label{fig:position-holding-acc}
    }
    \caption{Flight Test: position holding and scanning pattern maneuvers. The in-flight scanning dynamics are shown in (a), with position holding accuracy analysis during hovering and scanning maneuver in (b).}
\end{figure*}

At each location, the drone performs a scanning maneuver transitioning to hovering flight and changes orientation progressively to collect samples of all headings. The average position-holding accuracy is in the decimeter range and is provided by the accessory \gls{gnss} receiver, while the main receiver performs the scan. \cref{fig:flight-scanning} shows the scanning pattern and is representative of the maneuver performed at each scanning point.

Position-holding accuracy in hovering is shown in \cref{fig:position-holding-acc}, where the position holding at each heading is analyzed for one scanning location. One major limitation coming from the fixed-wing configuration is the sensitivity in positioning accuracy the \gls{vtol} can achieve when subject to cross-winds orthogonal to the wing plane. This effect is seen in the asymmetric error in the \gls{llh} frame and is hard to mitigate. Nevertheless, the error is in the order of tens of centimeters and does not influence the quality of the obtained scans. 

\begin{figure*}[]
    \captionsetup[subfloat]{farskip=2pt,captionskip=1pt}
    \setlength{\tabcolsep}{2pt} % Adjust spacing between columns if necessary
    \begin{tabular}{@{}p{0.5\textwidth}@{}p{0.5\textwidth}@{}}
        \centering
        \parbox{\linewidth}{%
            \centering
            \subfloat[]{
                \includegraphics[height=0.4\textwidth]{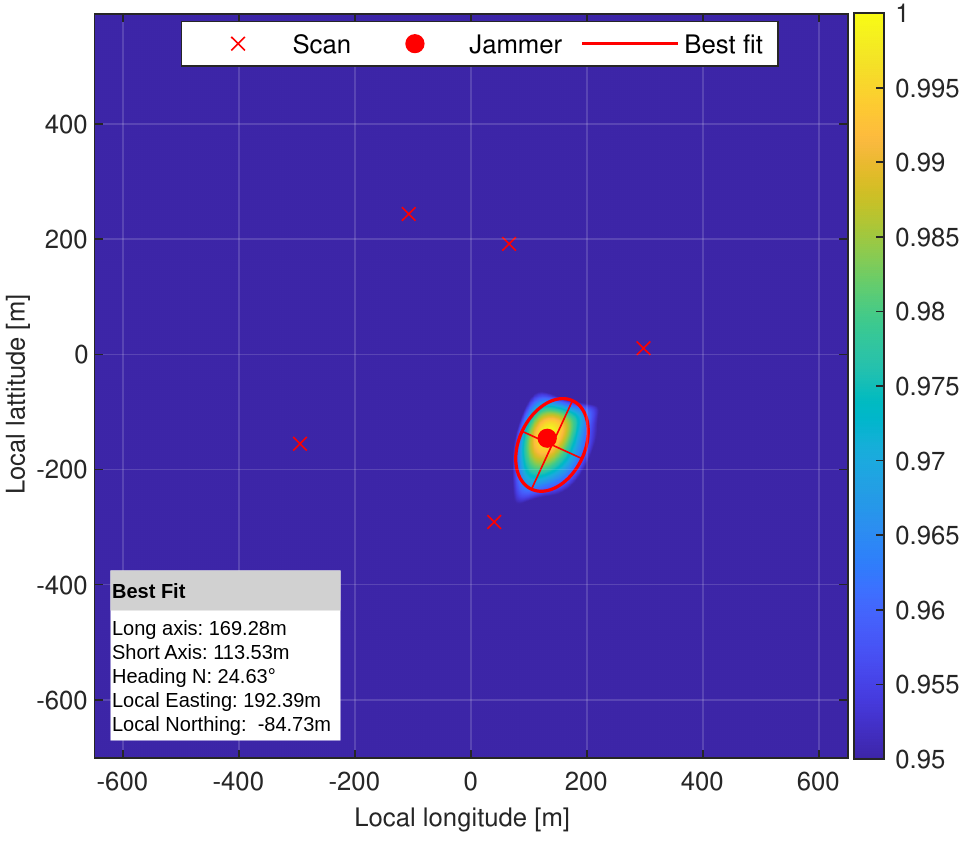} % Fixed height for consistency
                \label{fig:live-jammer-mapfusion}
            }
        }%
        &
        \parbox{\linewidth}{%
            \centering
            \subfloat[]{
                \includegraphics[height=0.4\textwidth]{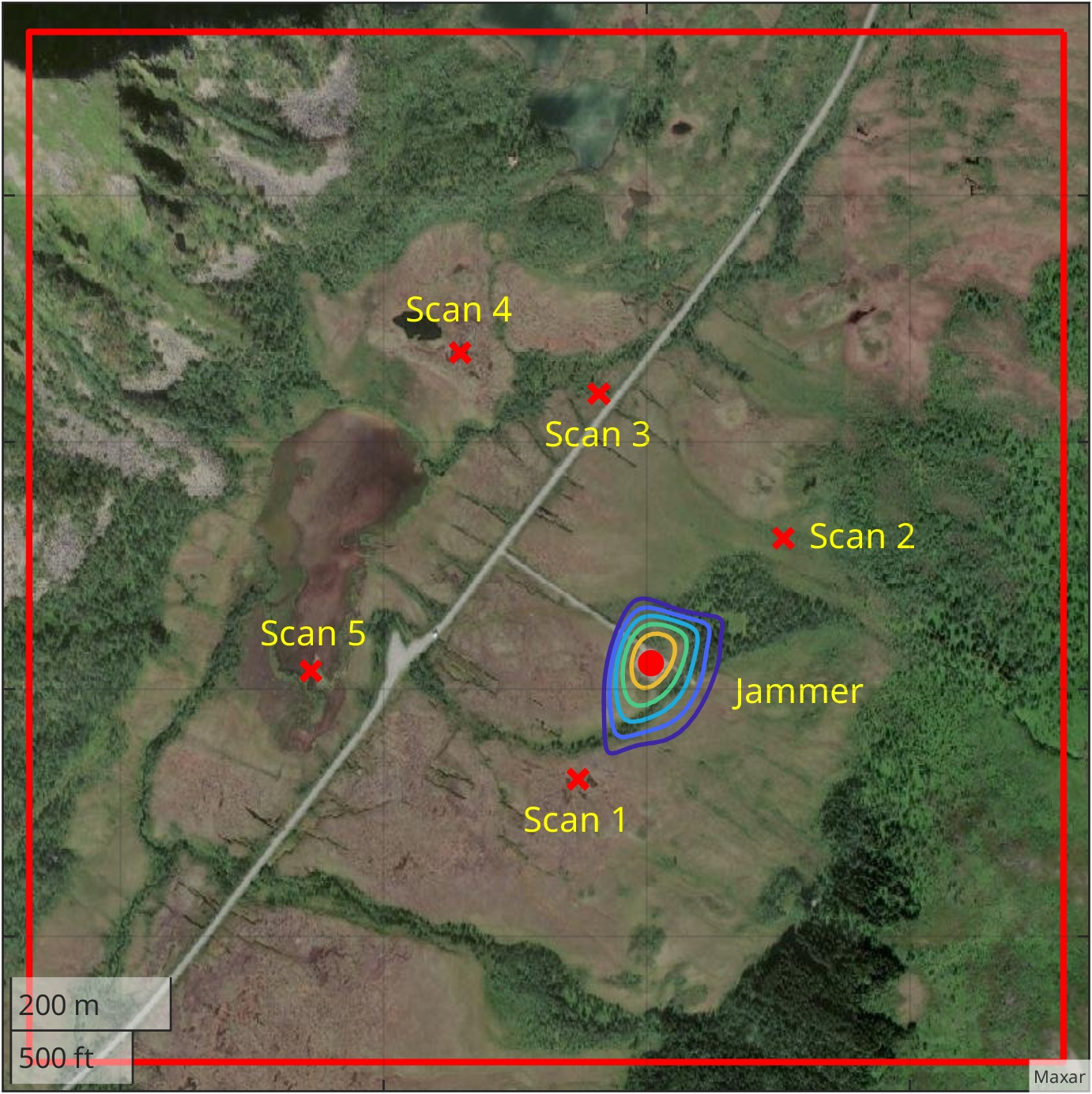} % Same height as left figure
                \label{fig:live-jammer-convergence}
            }
        }%
    \end{tabular}
    \caption{OTA\_Flight\_1: deployment of two \gls{ppd}s in the test area (at the same position) and successful localization based on long-distance measurements. Area center coordinates and dimensions refer to the relative coordinate system of the test area. The likelihood distribution of the jammer position and area of maximum degradation (a) and its convergence at the jammer location, in overlay with the satellite map of the test area (b).}
    \label{fig:ota-dyn-mapping-eval}
\end{figure*}

The localization of interference from \cref{algo:mapfusion} estimates the position of the jammers based on the samples collected at the front end. \cref{fig:live-jammer-mapfusion} shows the converged localization of the jammer, where the likelihood is rescaled to highlight the area where the jammer is more likely to be placed. Specifically, the scale is representative of the probability of a \gls{gnss} receiver entering such an area to be effectively jammed causing loss of \gls{pnt}. The extent of the contour is representative of the jammer area of coverage, which ultimately depends on the accuracy of the antenna heading during measurements. The best fit shown in \cref{fig:live-jammer-mapfusion} shows the extent and position of the centroid of the area of maximum effect of the adversarial transmitter.
Similarly, \cref{fig:live-jammer-convergence} shows the localization convergence around the jammer true position and given the favorable geometrical distribution of the scanning points the affected area is also mapped accurately. 

In this specific case, the survey points are distributed around the source of interference. This makes the localization precise and allows for the closure of the surfaces influenced by the transmitter. It is not always possible to achieve closure of the surface defining the affected area, for example, when the \gls{vtol} cannot fly all around the transmitter. This can be due to obstacles in the survey area or an extension of the jammer area of influence beyond the reach of the \gls{vtol} platform. Nevertheless, the mapping algorithm will highlight areas where the adversary is less effective allowing the \gls{vtol} to safely transit past the interference source.

\begin{figure*}
    \centering
    \begin{tabular}{@{}p{0.25\textwidth}@{}p{0.25\textwidth}@{}p{0.25\textwidth}@{}p{0.25\textwidth}@{}}
    \subfloat[]{
        \includegraphics[width=\linewidth]{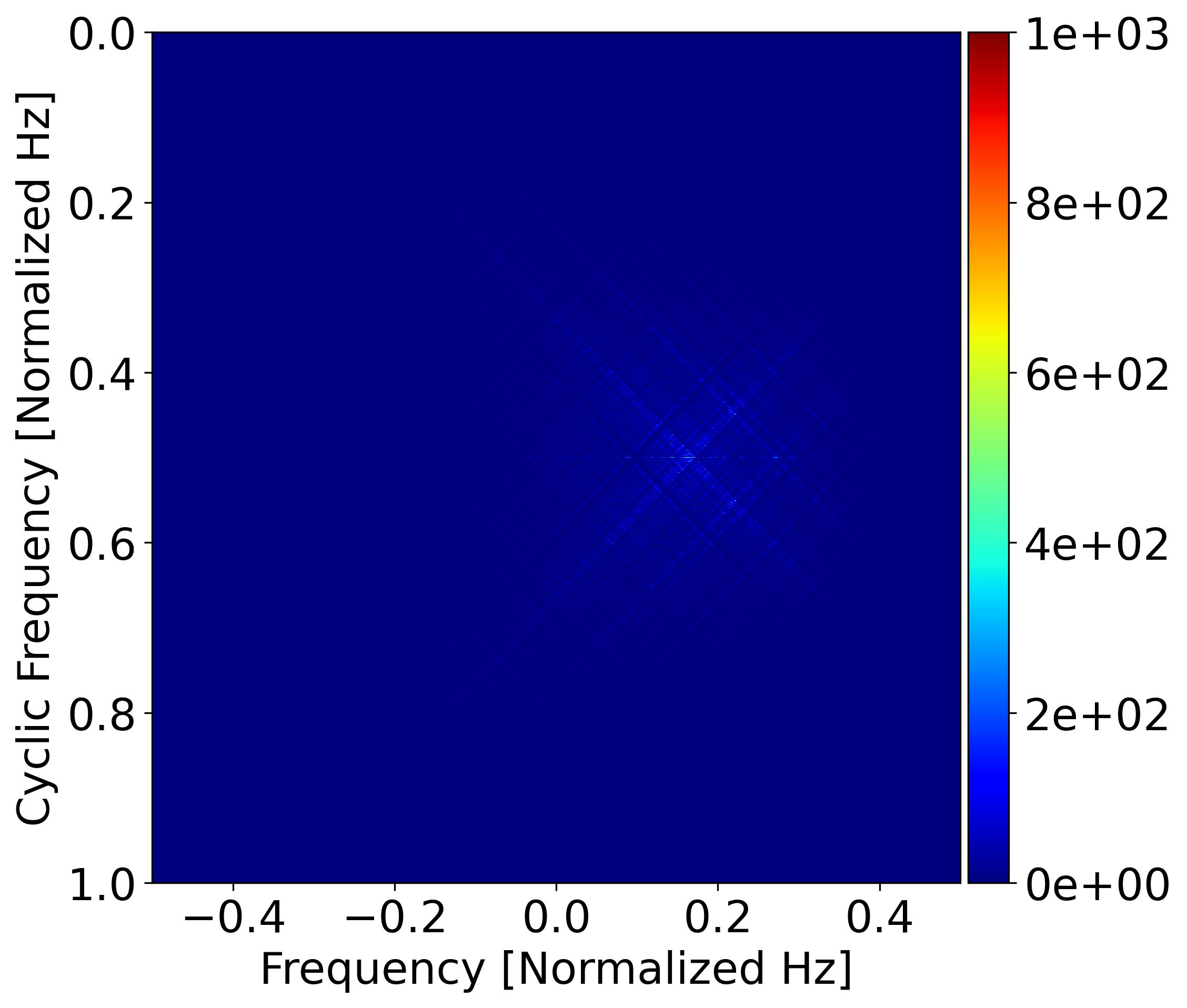}
        \label{fig:flight-benign-1}
    }&
    \subfloat[]{
        \includegraphics[width=\linewidth]{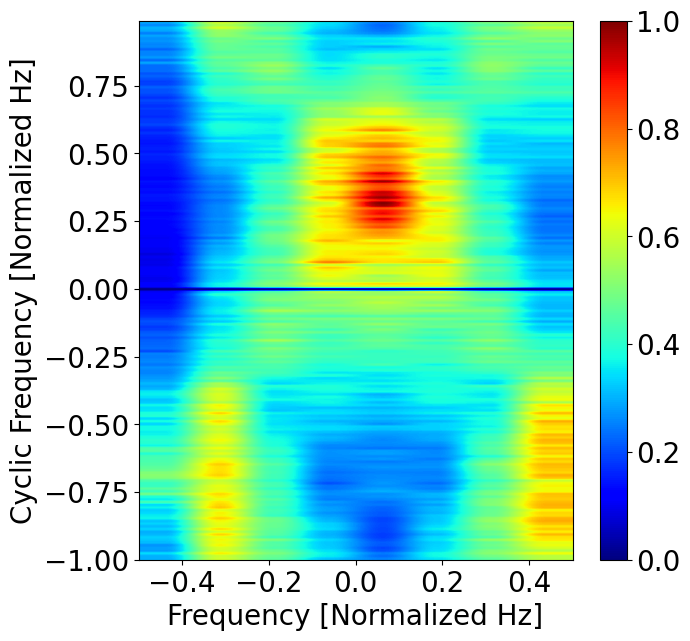}
        \label{fig:flight-benign-1-coh}
    }&
    \subfloat[]{
        \includegraphics[width=\linewidth]{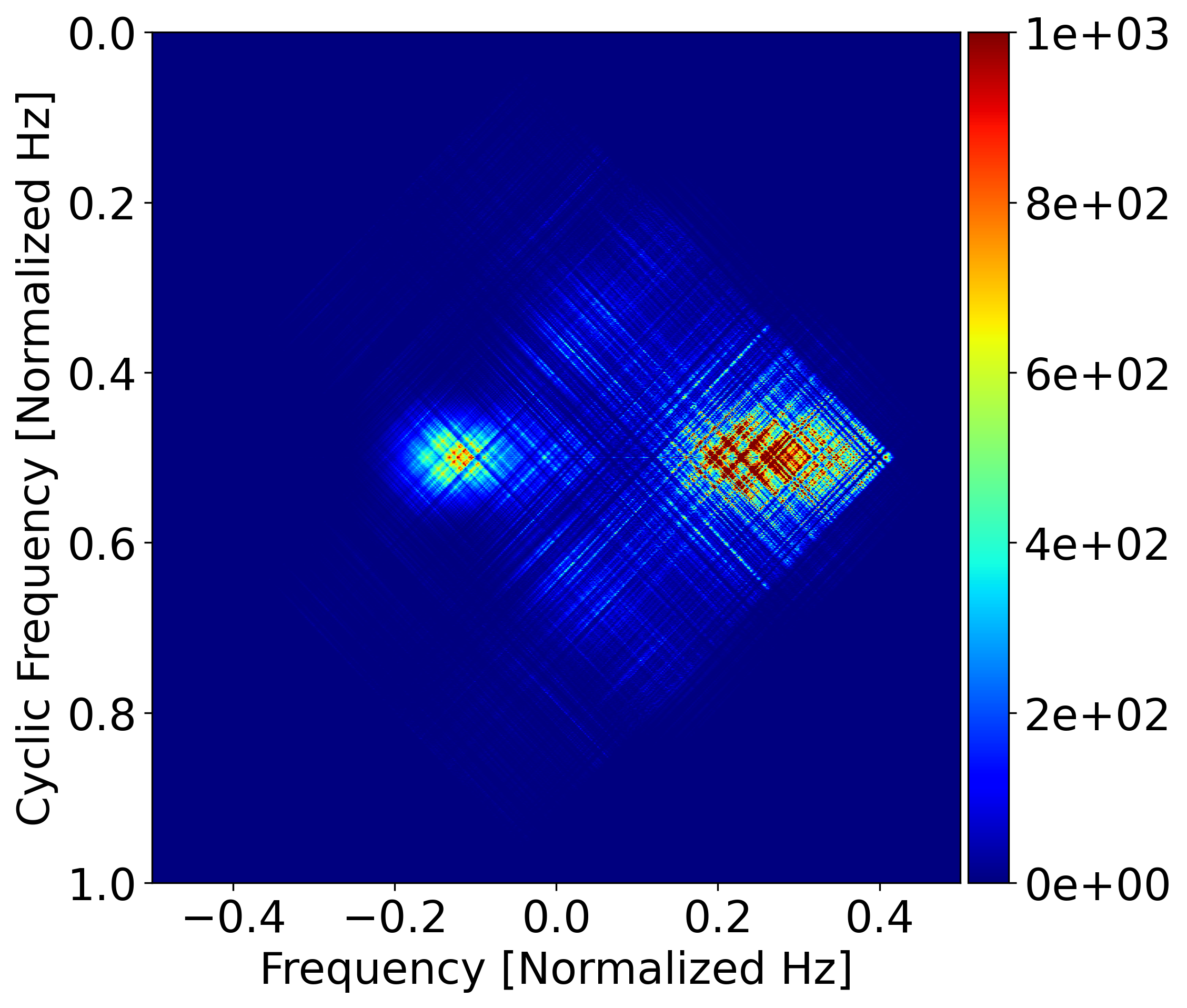}
        \label{fig:flight-jammer-weak}
    }&
    \subfloat[]{
        \includegraphics[width=\linewidth]{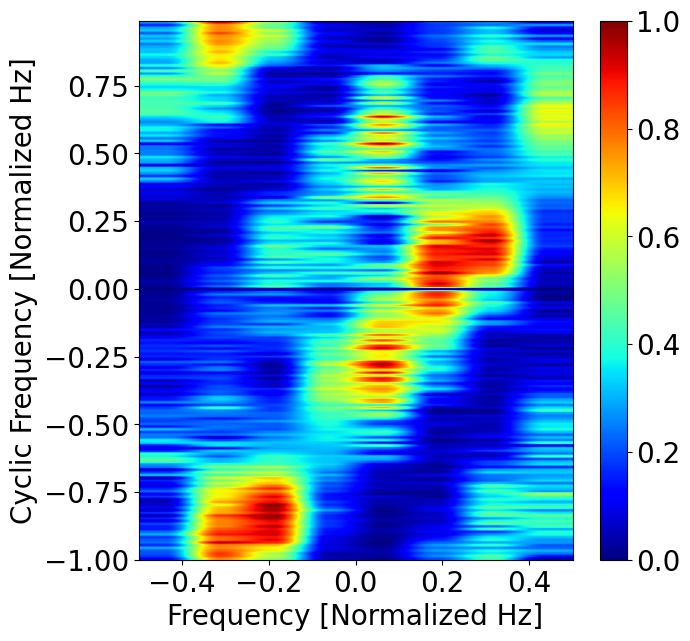}
        \label{fig:flight-jammer-weak-coh}
    }
    \end{tabular}
    \caption{OTA\_Flight\_1\_a: Limited interference is measured and the \gls{vtol} performs as expected. Still, even with a minimal level of interference the structure of the \gls{scd} gives information regarding a degraded situation. Conditions without significant interference (a) and spectral coherence of signal in the benign case (b); approaching the jammer: acceptable level of interference and unaffected navigation (c) with \gls{coh} in case of limited interference showing a sweeping jammer in the distance (d).}
    \label{fig:ota-dyn-benign-eval}
\end{figure*}

In \cref{fig:flight-benign-1} the drone starts a survey, following the trajectory from \cref{fig:live-jammer-flight}. When the \gls{vtol} points the main lobe of the scanning antenna towards an area without adversarial transmitters, the detector shows no excess power is present the \gls{scd} is overall unstructured, with a low relative auto-correlation power as in \cref{fig:flight-benign-1,fig:flight-benign-1-coh}. Later in the flight, the \gls{vtol} approaches an area with increased jamming power as seen in \cref{fig:flight-jammer-weak} where the spectral structures in the \gls{scd} are significantly more defined but still navigation is possible. The presence of structures in the \gls{coh} in \cref{fig:flight-jammer-weak-coh} indicates a nearby jammer. %-- ref to table and the measured jammer

\begin{figure*}
    \centering
    \begin{tabular}{@{}p{0.25\textwidth}@{}p{0.25\textwidth}@{}p{0.25\textwidth}@{}p{0.25\textwidth}@{}}
    \subfloat[]{
        \includegraphics[width=\linewidth]{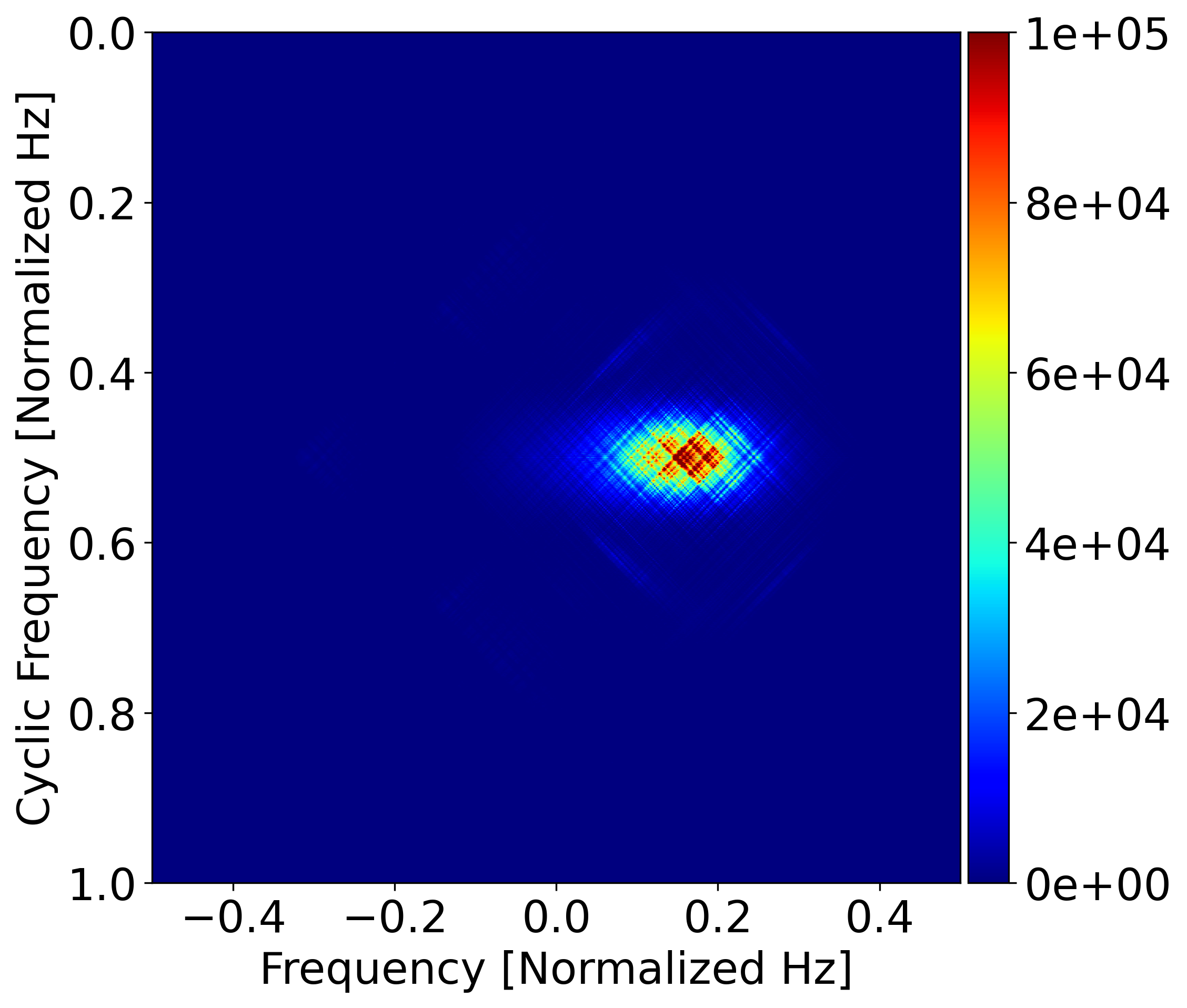}
        \label{fig:flight-jammer-strong}
    }&
    \subfloat[]{
        \includegraphics[width=\linewidth]{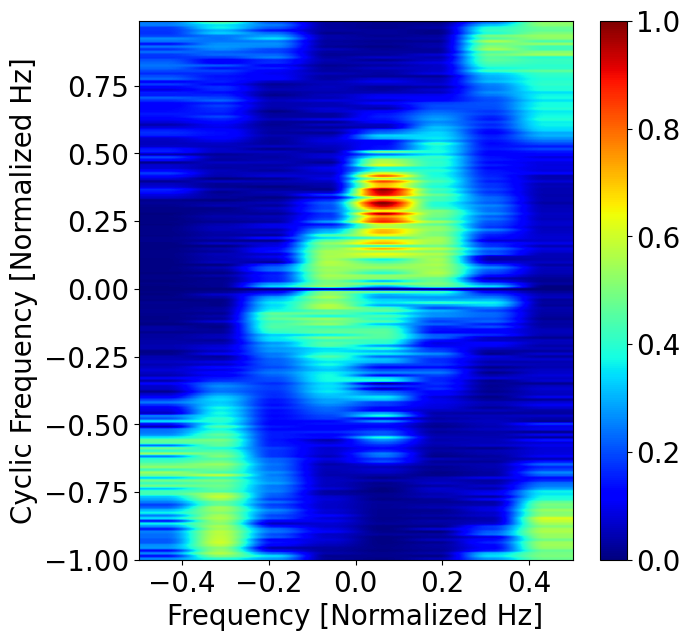}
        \label{fig:flight-jammer-strong-coh}
    }&
    \subfloat[]{
        \includegraphics[width=\linewidth]{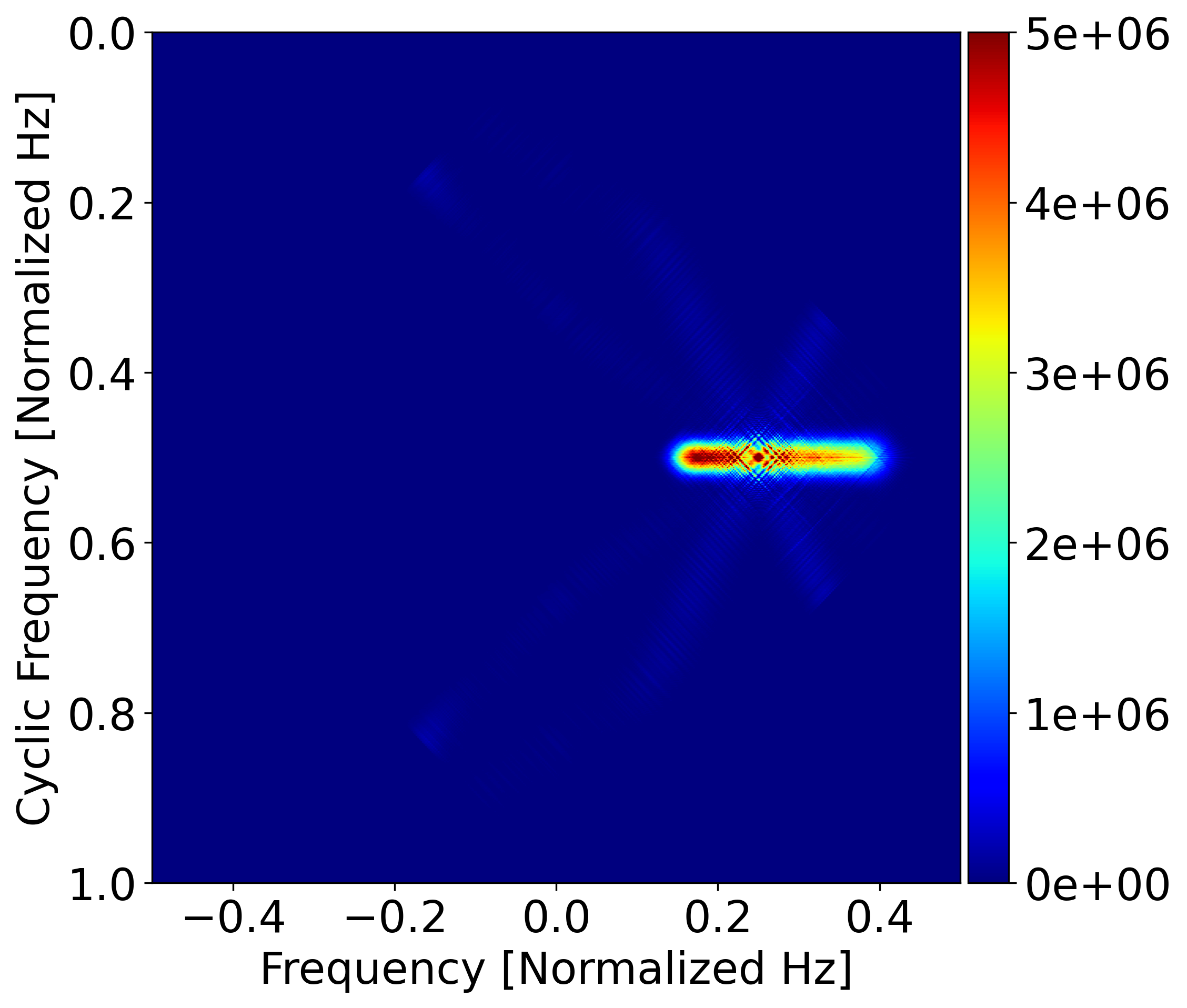}
        \label{fig:flight-jammer-strong-2}
    }&
    \subfloat[]{
        \includegraphics[width=\linewidth]{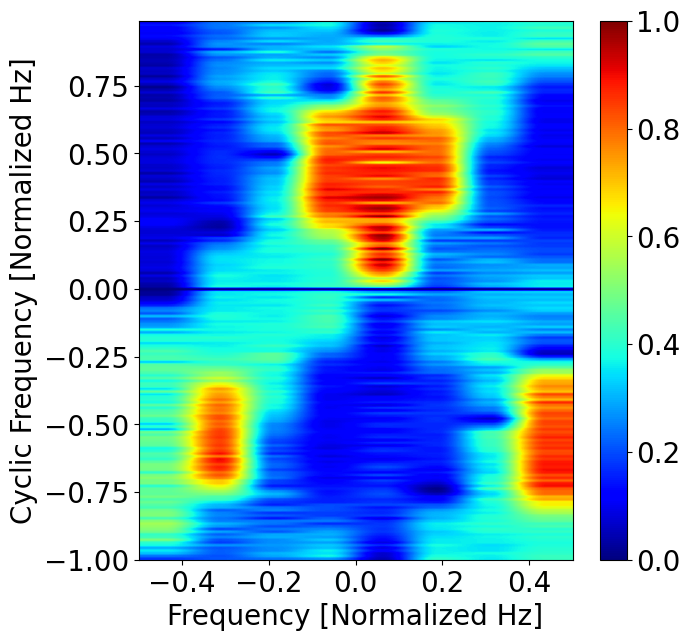}
        \label{fig:flight-jammer-strong-2-coh}
    }%
    \end{tabular}
    \caption{OTA\_Flight\_1\_b: High interference detection where navigation accuracy is reduced within the effective zone of the jammer. Different jammers are identified based on the \gls{scd} structure. LP\_Swept: strong swept narrow-band jammer (a) and \gls{coh} of fast the aggressive swept jammer (b); LP\_Multi\_1: Strong wideband jammer (c) and \gls{coh} of the wideband multiband jammer (d).}
    \label{fig:ota-dyn-hard-eval}
\end{figure*}

In \cref{fig:flight-jammer-strong,fig:flight-jammer-strong-2} the \gls{vtol} antenna is directly facing several powerful jammers present at the test location.
In this case, the two jammers are a chirp jammer of \SI{15}{\mega\hertz} bandwidth and a swept L1 jammer. This is made more evident in \cref{fig:flight-jammer-strong-2-coh}, where a superposition of multiple structures is measured. A chirp jammer is clear at L1 center frequency while a swept jammer crosses the entire monitored bandwidth and a narrow-band tone. In \cref{fig:flight-jammer-strong-coh}, the chirp jammer is less effective, but the swept jammer is still present.

Navigation in both cases would be severely impaired and would be unreliable in the areas affected by the jammer. This shows the effectiveness of such a method: leveraging high gain and directivity of the antenna, the \gls{vtol} never enters the affected zone, as shown in \cref{fig:live-jammer-mapfusion}. By sampling the RF spectrum outside the denied area, the \gls{vtol} is never in a condition like the one shown in \cref{fig:degradation-static} where navigation is denied.  

\textcolor{black}{One remark is important here. As the jammer power is potentially unknown and the receiver uses adaptive gain control to maximize the ADC range, the absolute power level estimation of the transmitter is complex if not unfeasible unless a calibrated front-end and antenna pair is used. While possible, this is beyond the scope presented here of using only the receiver available in the VTOL platform. Nevertheless, the relative power estimate is enough to detect the presence of the jammer. This highlights one of the advantages of combining raw IQ samples from the receiver and cyclostationary processing, as the relative power/frequency estimates are effective in detecting the jammer, without requiring complex acquisition systems.}

\subsection{Automatic detection and identification of interference}
\textcolor{black}{
From the results shown in \cref{fig:ota-dyn-benign-eval,fig:ota-dyn-hard-eval}, the detection of the jammer can be extended from a simple energy detector as in \cref{eq:psd}, to a more complex model. The first and most straightforward way is integrating the \gls{scd} values for $\alpha=0$. Any significant power from a strong noise transmitter would result in a strong autocorrelation of the jamming signal, but this is practically, a spectral power detector. We recall here that, in the protected navigation bands the total received power from any point on the ground should be comparable in magnitude to the thermal noise at the front-end. Any power level received, even if low enough to not constitute a navigation risk can be considered as potentially adversarial.
}

\begin{figure}
    \centering
    \includegraphics[width=0.85\linewidth]{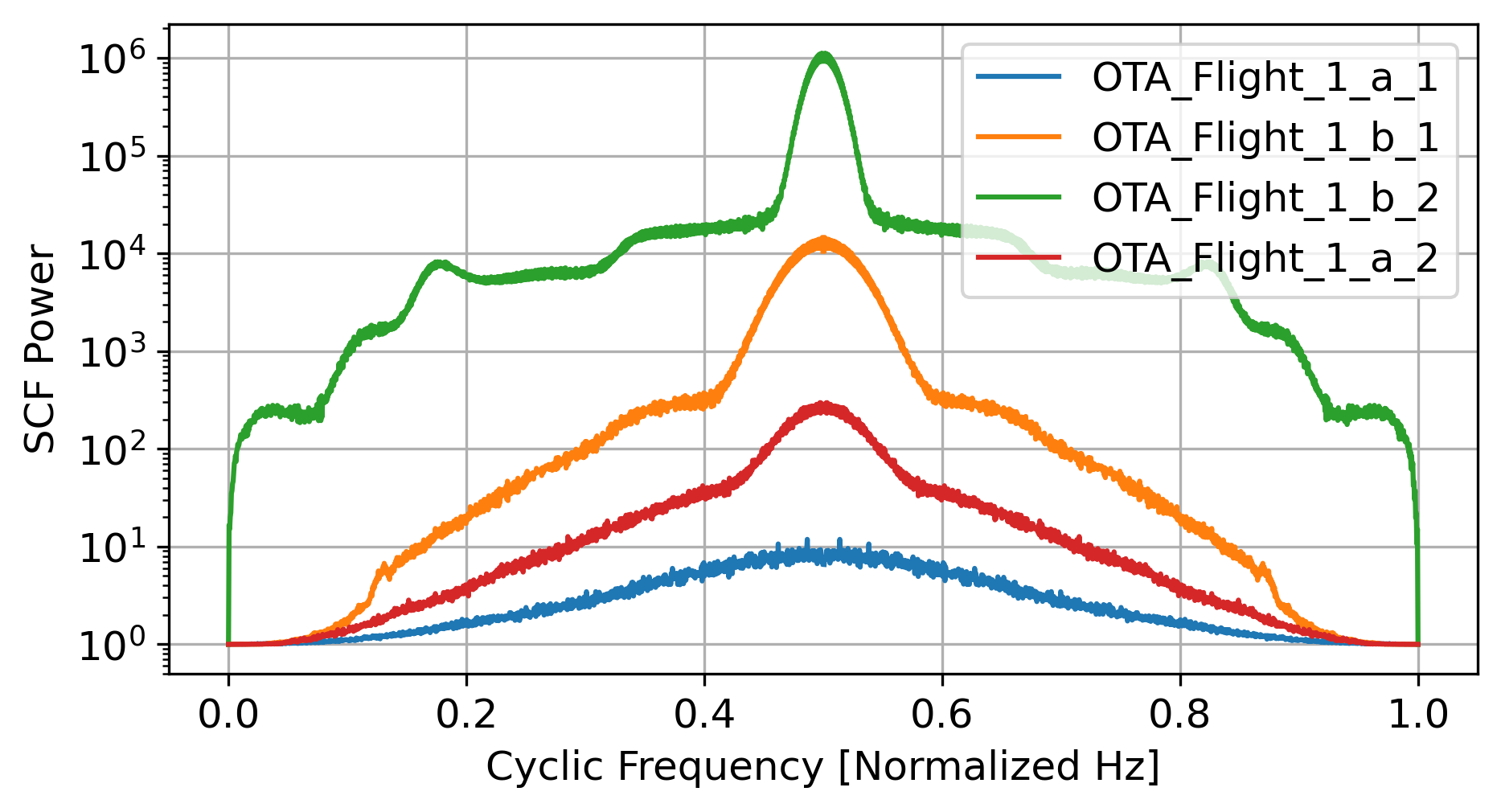}
    \caption{Flight tests compound 1D collapsed SCF based on \cref{eq:detection-scf}.}
    \label{fig:compound-scf-flight}
\end{figure}

\textcolor{black}{
From the \gls{scd} we can estimate the presence of the jammer by analyzing the magnitude of the \gls{scd} itself. Integration across the cyclic frequencies to obtain the so-called $\alpha$-profile is used to detect at the same time the presence of a jammer and its cyclic behavior. \cref{fig:compound-scf-flight} shows exactly this behavior. By collapsing the \gls{scd} obtained from the \gls{fam} calculation along the cyclostationary frequencies as described in \cref{eq:detection-scf}, we can detect the power of cyclostationary components as a whole. 
The case of \cref{fig:flight-benign-1}, where no significant interference is detected leads to a flat relative power where the only contribution is due to AWGN nise at the fron-end. On the other hand, \cref{fig:flight-jammer-strong-2} leads to a higher power.
A threshold detector is used to evaluate the magnitude of the cyclostationary peaks and detect the presence of a jammer. This threshold is related to the benign case and can be calculated relatively to it. The maximum amplitude of the benign case represents the detection threshold, any power peak above such maximum value can be considered adversarial with an effect directly proportional to the measured magnitude.
}

\begin{figure}
    \centering
    \includegraphics[width=0.85\linewidth]{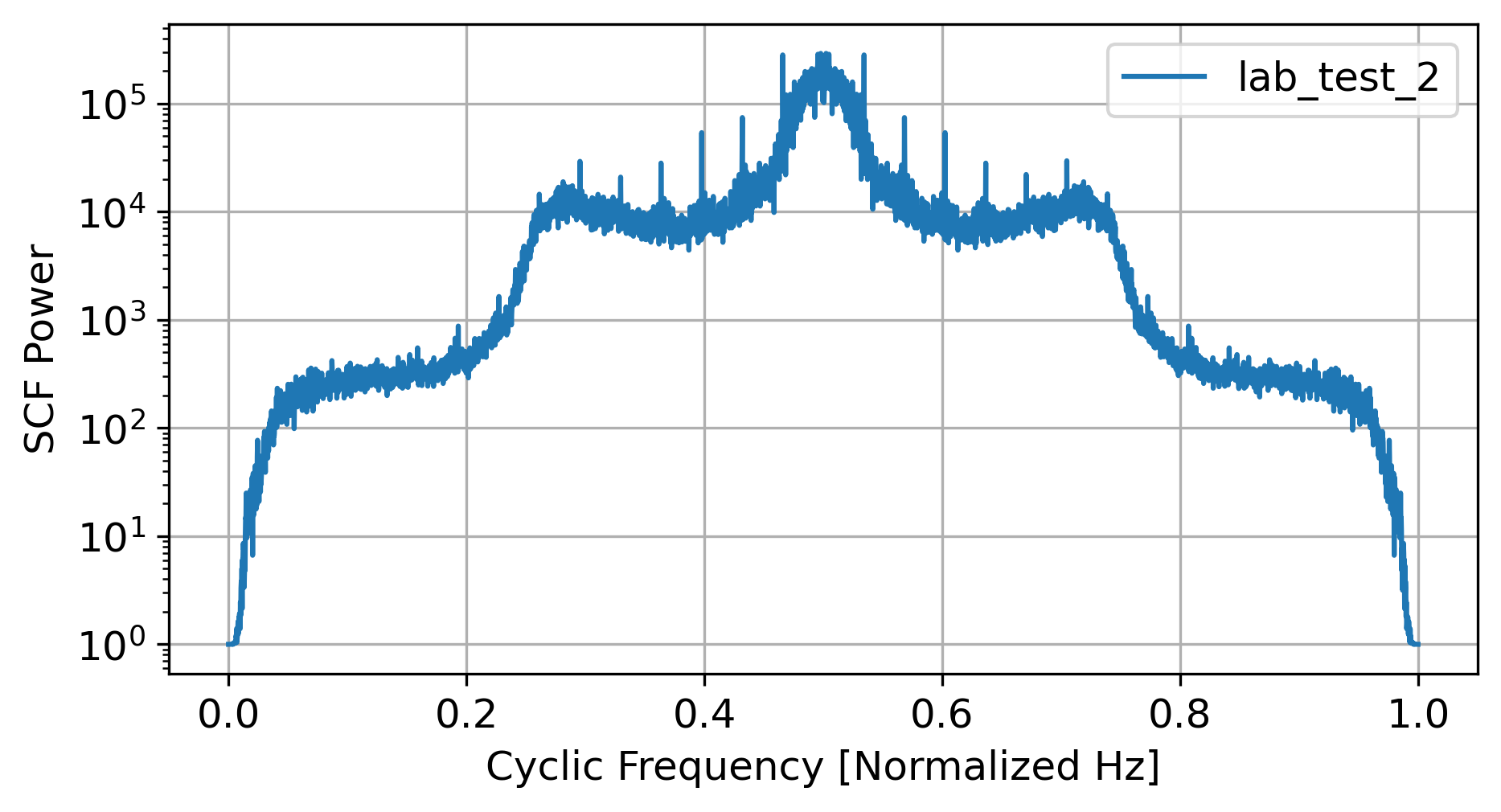}
    \caption{BPSK jammer compound 1D collapsed SCF based on \cref{eq:detection-scf}.}
    \label{fig:compound-scf-flight-bpsk}
\end{figure}

Similarly, the evaluation of the same metric on a signal with a more complex frequency content shows that also the cyclostationary frequencies are seen in the integrated plot, as expected. This is shown in \cref{fig:compound-scf-flight-bpsk}, based on data collected during jamming using a PRN-based signal.

\textcolor{black}{From the tests performed, a high value of the spectral correlation is directly linked with the presence of an adversarial transmitter. Several orders of magnitude differences in the \gls{scd} peaks amplitude can be seen comparing \cref{fig:flight-benign-1} and \cref{fig:flight-jammer-strong}. \textcolor{black}{This gives evidence of the presence of the jammer, and compared to the spectral power density, it also provides insight into the frequency/time characteristics of the jammer, such as the frequencies of the cyclostationary modes and the modulation rate in the case of, i.e. the BPSK case.}}

\textcolor{black}{
Additionally, the \gls{scd} analysis can be subdivided into clusters for specific cyclic and carrier frequencies, limiting the \gls{scd} power/frequency analysis to those particular areas of interest. This is an optimization compared to analyzing the entire \gls{scd} as it only operates on fewer samples around the area of interest. In this way, interference that might be detected but is not directly located at the frequencies of interest can be ignored. Such mask information can be extracted by measuring the receiver front-end input filter response, but this requires specific adaptation for each receiver.}

\textcolor{black}{Alternatively, it can also be obtained as the reciprocal of the modulation power to the receiver acquisition noise floor. Such value can be estimated based on the modulation properties of each signal and offset by the receiver sensitivity. For example, a template \gls{scd} for each \gls{gnss} legitimate signal can be pre-computed (given that each modulation is known), and its complementary power to the receiver acquisition threshold used as a power threshold for the \gls{scd}. In such case, for all experiments shown here, the detector would trigger based on abnormal power at cyclic frequencies not justified by the legitimate signal modulation. A reasonable threshold for the detection can be extracted from the lowest peak in \cref{fig:compound-scf-flight} (corresponding to the benign case, OTA\_Flight\_1\_a\_1).}

\textcolor{black}{
Ultimately, the two representations should be evaluated jointly. The \gls{scd} analysis is useful to estimate the actual strength of the jammer with some insight of the frequency component. The \gls{coh} on the other hand provides specific insight into the jammer modulation, bandwidth, and frequency content useful for identification purposes. As a normalized representation of the \gls{scd}, the \gls{coh} is effective in showing the power of the cyclostationary components relative to each other, within the same signal, allowing to characterize multiple overlapping signals produced by the same or independent transmitters. 
}

\textcolor{black}{
Given the structure of the resulting \gls{scd} and \gls{coh}, it is possible to use this information to train an automatic classifier capable of identifying the transmitter. As discussed in \cref{sec:related_work}, such an approach comes at a non-significant challenge, requiring an exhaustive training set, covering many possible transmitters. Nevertheless, compared to the existing methods that rely on spectrograms, spectral correlation representations allow for a more compact and structured approach. This is due to the higher amount of information available. In this case, not only is the power and spectral content available, but the classifier also has access to information concerning the periodicity of the signal.
}

\textcolor{black}{
One problematic aspect of this approach, as mentioned, is the requirement of a comprehensive database of signal structures. On the other hand, the \gls{coh} analysis allows for a simpler implementation of the database which can be created based on the theoretical expression of the most common civilian jamming signals as the dominant features in the measured signal and the simulated one are the same. This approach would benefit from the well-defined structure of the spectral correlations. Nevertheless, the investigation of the actual effectiveness and influence of other parameters such as sampling frequency and front-end bandwidth is left for future work.
}

\subsection{Long range estimation of the GNSS degradation}

\textcolor{black}{Estimation of the navigation quality in a distant area is a complex problem, in particular when it is performed without being actively affected by the jammer. Despite this, it is of interest as it allows real-time decisions regarding the chances the receiver has in surviving a specific attack. \textcolor{black}{Given the anti-jamming capabilities of modern high-performance receivers, attacks like sweeping tones are likely to be addressed with advanced filtering, but this is not guaranteed to avoid degradation of the PNT solution. From experimental evidence during testing, more complex modulations, like aggressive PRN-like transmitters are difficult to address at the receiver. Based on the \gls{coh} a receiver can estimate the nature of the signal.}}
The degradation of the \gls{gnss} \gls{pnt} quality usually depends on the ability of the jammer to disrupt the channel, the receiver's robustness and adaptiveness in counteracting the attacker, and the level of accepted degradation. It is worth noticing that when collecting samples at a distance, in not all cases the \gls{gnss} receiver was denied a \gls{pnt} solution. Nevertheless, the reduction in accuracy (measured as standard deviations of position estimation error in the horizontal and vertical dimension, EPH, and EPV) was considerable. As our method relies on pre-correlation samples, it is not bound to the latency of the adaptation of the internal \gls{gnss} \gls{pnt} engine. This makes the detection and identification faster compared to \gls{sqm} countermeasures, but a quantitative analysis of this information is left for future investigation and it is likely to be receiver-dependant.

\section{Conclusions}
\label{sec:conclusion}
We presented an extended method for the localization of adversarial transmitters by an aerial platform and an application of cyclostationary signal analysis to IQ samples provided by a consumer \gls{gnss} receiver. The method is capable of detecting not only the presence of the jammer but also the specific modulation used and extracting for the jammed cases the time-domain properties of the signal. The presented implementation and application show a more powerful tool compared to direct analysis of the \gls{pnt} engine solution or its availability, even when \gls{sqm} countermeasures are in place. Combined with an airborne platform capable of precise and advanced maneuvering, the system is effective in localizing sources of interference at long distances, without necessarily being affected by the adversarial transmitter. \textcolor{black}{The jammer detector based on \gls{scd}/\gls{coh} performs as well as with an \gls{sdr} when using samples provided by a \gls{gnss} receiver, as long as the provided samples are raw baseband IQ values.} Additionally, the lower sample rate and the snapshot-based spectrum view allow for processing even in lower-performance systems in contrast with \gls{sdr} streaming that requires high post-processing performance. \textcolor{black}{Future development could enable the aerial platform to autonomously perform online detection and identification. This would allow feedback of the resulting spectral awareness assessments directly to the flight controller, making the autonomous platform not only able to navigate unattended but also to avoid areas with degraded \gls{pnt}.} This overall increases both the robustness and the safety of the unmanned system.

The detection is successful in all cases tested and provides valuable information even before the complete loss of \gls{pnt}, as shown in the preliminary laboratory evaluation and the static tests. This is advantageous for early detection of interference in highly mobile scenarios, as shown in the tests performed with active flight. Identification of different types of jammers was achieved in all scenarios, with a clear distinction between modulations, sweep, and spectral content.

The main challenge of deploying automatic spectral awareness systems is the classification of the signal. While the coarse approach based on the shape of the cross-correlation is effective, a pattern recognition approach is more effective. Establishing a library of possible patterns is challenging, especially when the degrees of freedom in signal design at the attacker's disposal are many: it would be necessary to include in the training model many different signals with different properties. This is a challenging scenario that is currently under development and would improve the performance of the system even further. 

\bibliographystyle{IEEEtran}
\bibliography{sample_full}

%% This biostyle allows you to insert your photo size 1in X 1.25in
\begin{IEEEbiography}[{\includegraphics[width=1in,height=1.25in,clip,keepaspectratio]{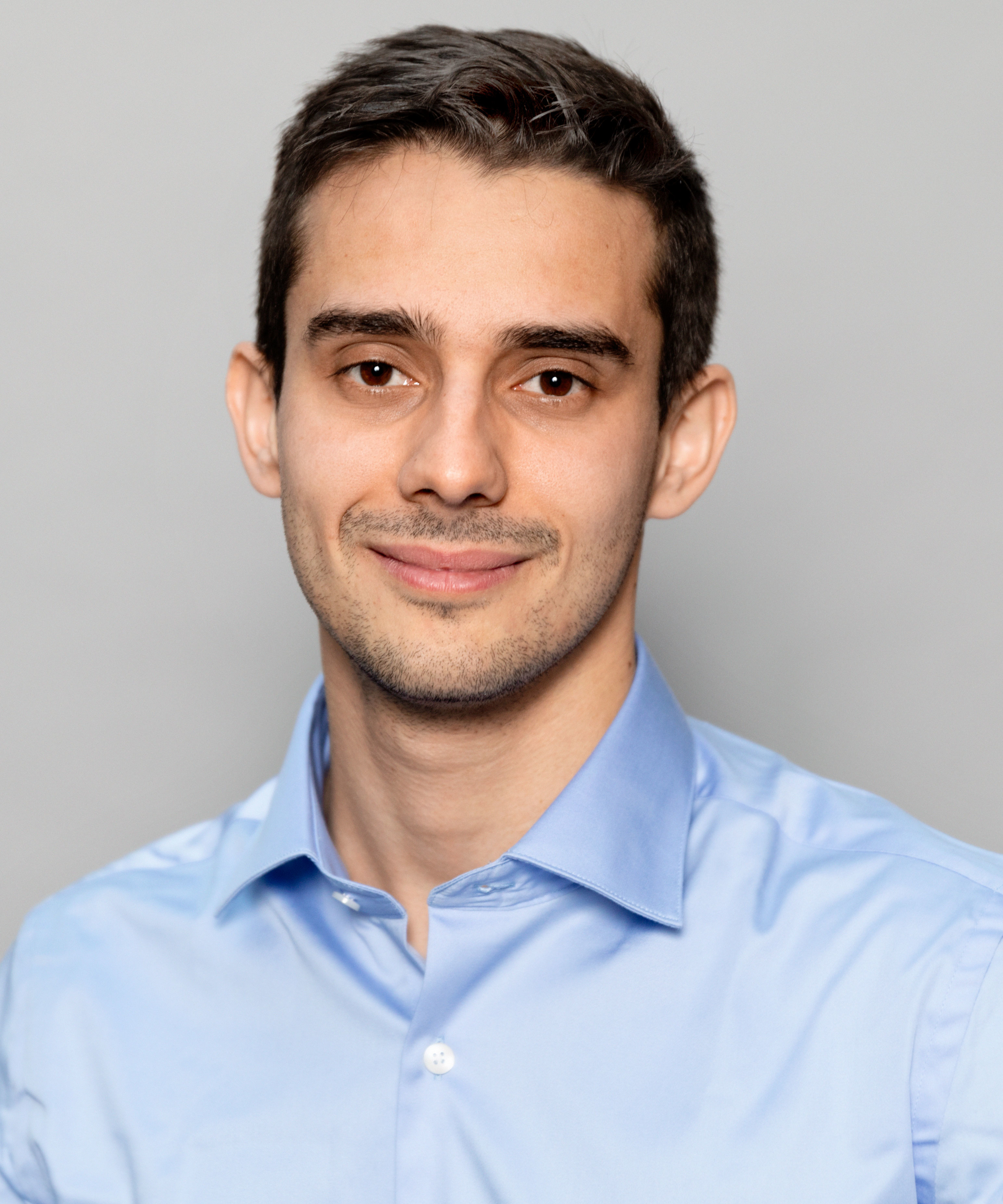}}]{Marco Spanghero} received his B.S. from Politecnico of Milano and an MSc degree from KTH Royal Institute of Technology, Stockholm, Sweden. He is currently a Ph.D. candidate with the Networked Systems Security (NSS) group at KTH, Stockholm, Sweden, and associate with the WASP program from the Knut and Alice Wallenberg Foundation. 
\end{IEEEbiography} 
\begin{IEEEbiography}[{\includegraphics[width=1in,height=1.25in,clip,keepaspectratio]{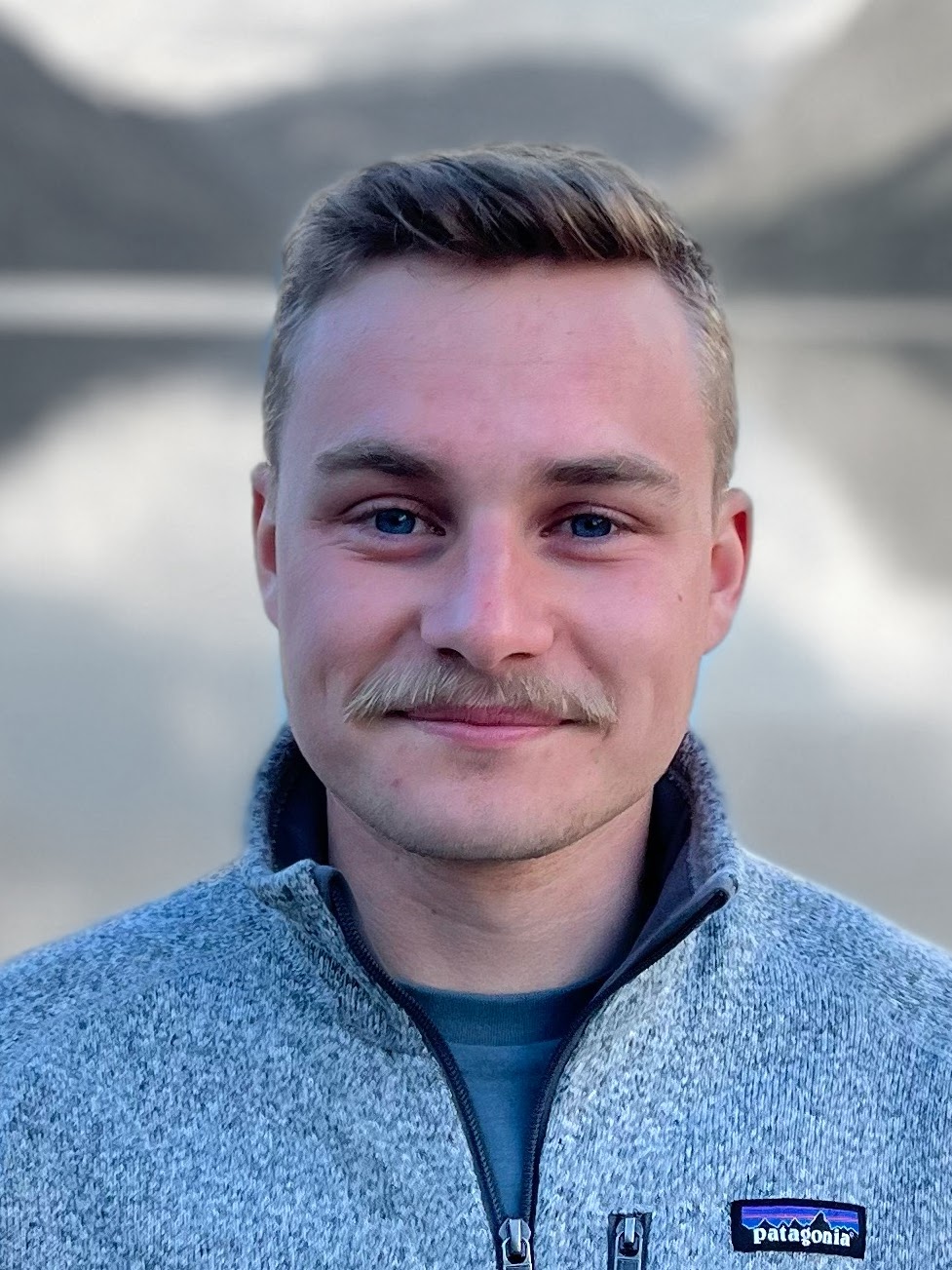}}]{Filip Geib}
earned his B.C. degree in Cybernetics and Robotics from the Czech Technical University in Prague in 2021. He furthered his education with an M.Sc. degree in Systems, Controls, and Robotics from the KTH Royal Institute of Technology, Stockholm, in 2023. At the time of this work, he contributed as a Robotics Software Engineer at Wingtra AG. 
\end{IEEEbiography}
\begin{IEEEbiography}[{\includegraphics[width=1in,height=1.25in,clip,keepaspectratio]{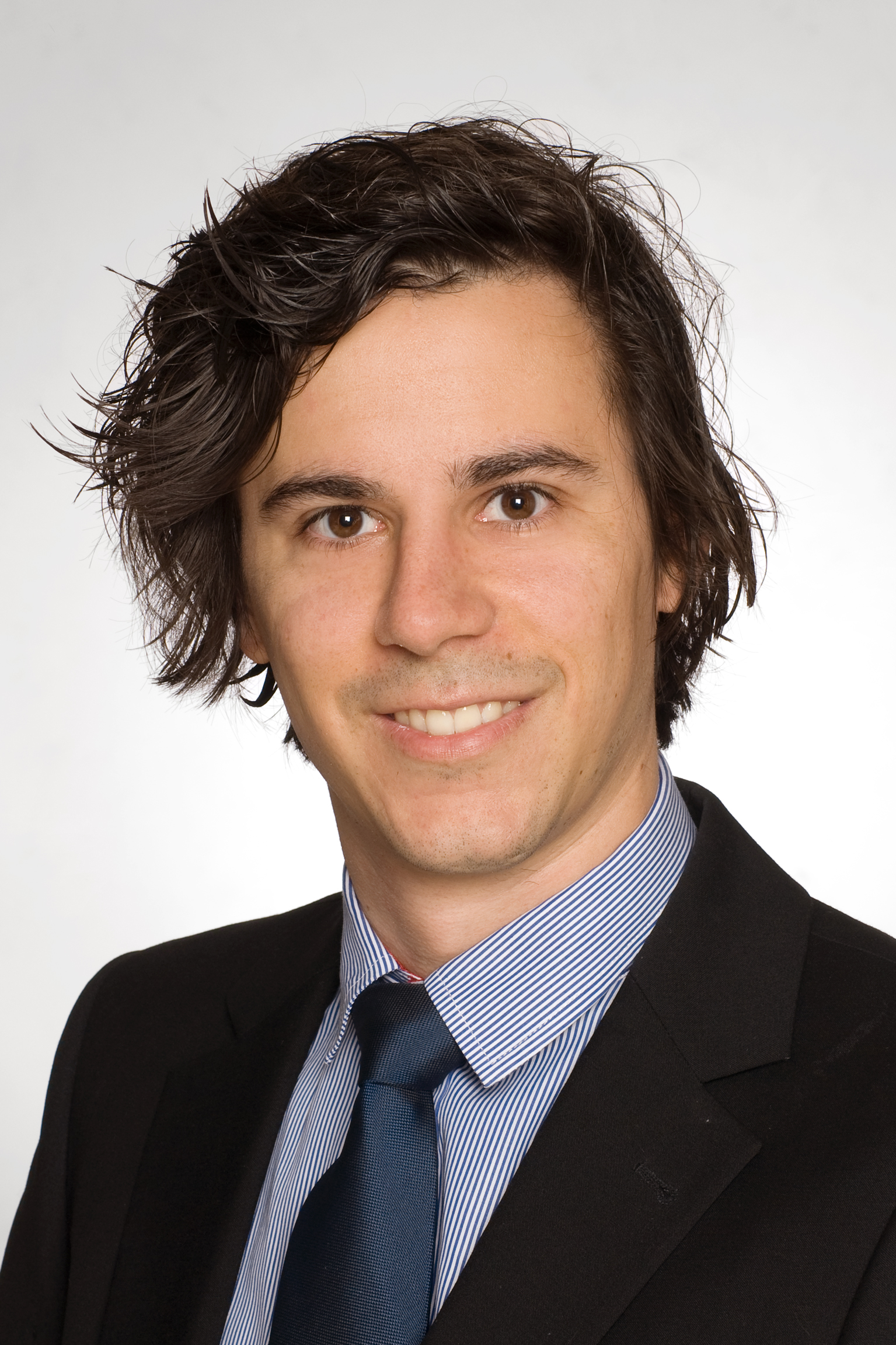}}]{Ronny Pannier}
received his MSc degree in Mechanical Engineering in 2015 from ETH Zurich in Switzerland. At the time of this work he contributed as a team-lead Systems Engineer at Wingtra AG. 
\end{IEEEbiography}
\begin{IEEEbiography}[{\includegraphics[width=1in,height=1.25in,clip,keepaspectratio]{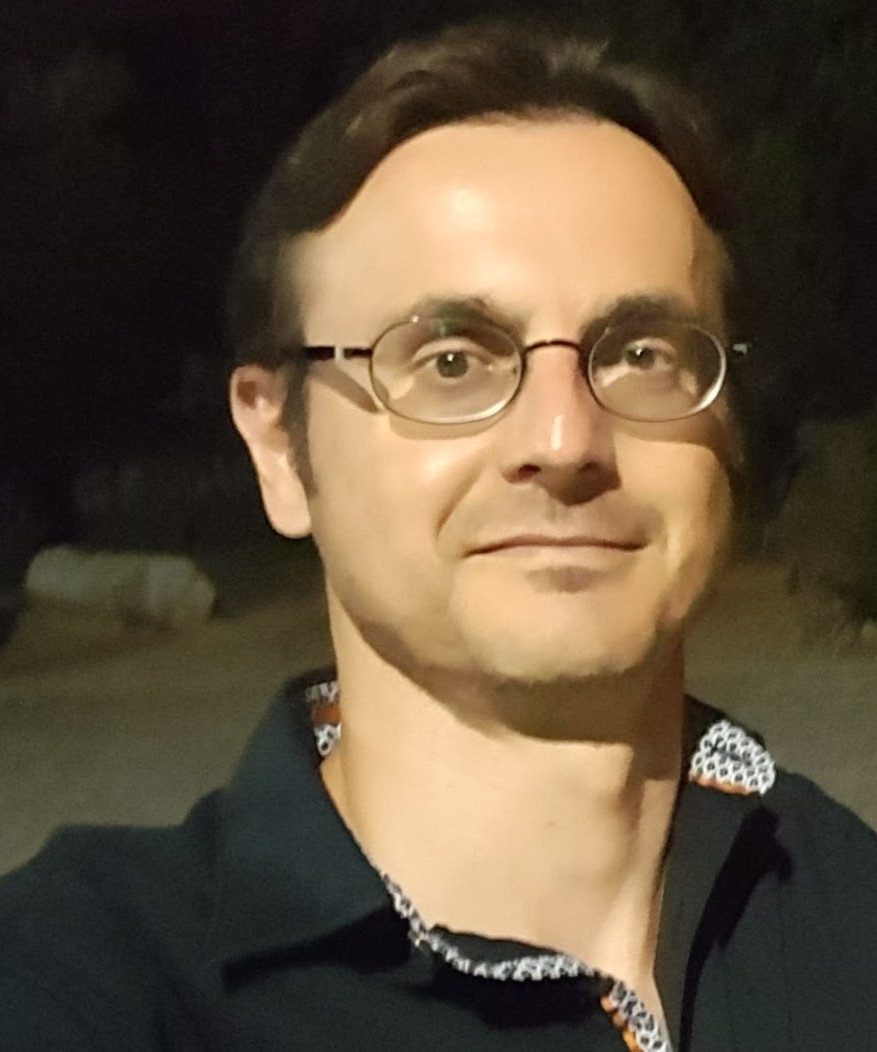}}]{Panos Papadimitratos}
(Fellow, IEEE) earned his Ph.D. degree from Cornell University, Ithaca, NY, USA. At KTH, Stockholm, Sweden, he leads the Networked Systems Security (NSS) group and he is a member of the Steering Committee of the Security Link Center. He serves or served as a member of the ACM WiSec and CANS conference steering committees and the PETS Editorial and Advisory Boards; Program Chair for the ACM WiSec’16, TRUST’16, and CANS’18 conferences; General Chair for the ACM WISec’18, PETS’19, and IEEE EuroS\&P’19 conferences; Associate Editor of the IEEE TMC, IEEE/ACM ToN and IET IFS journals, and Chair of the Caspar Bowden PET Award. Panos is a Fellow of the Young Academy of Europe, a Knut and Alice Wallenberg Academy Fellow, and an ACM Distinguished Member. NSS webpage: \url{https://www.eecs.kth.se/nss}.
\end{IEEEbiography}

\end{document}